\documentclass[11pt,a4paper,american]{extarticle}
\usepackage[T1]{fontenc}
\usepackage[latin1]{inputenc}
\usepackage{amsmath}
\usepackage{amssymb}
\usepackage{setspace}
\usepackage{esint}
\pdfoutput=1
\makeatletter

\usepackage{babel}
\usepackage{bbm}
\usepackage{hyperref}
\usepackage{authblk}
\usepackage{graphicx}
\usepackage{color}
\date{}
\allowdisplaybreaks
\numberwithin{equation}{section}
\author{Renann Lipinski Jusinskas\thanks{renannlj@fzu.cz}}
\affil{Institute of Physics of the Czech Academy of Sciences \\ CEICO - Central European Institute for Cosmology and Fundamental Physics
\authorcr  Na Slovance 2, 182 21, Prague - Czech Republic}

\makeatother

\usepackage{babel}
\begin{document}
\title{Chiral strings, the sectorized description and their integrated vertex
operators}
\maketitle
\begin{abstract}
A chiral string can be seen as an ordinary string in a singular gauge
for the worldsheet metric and has the ambitwistor string as its tensionless
limit. As proposed by Siegel, there is a one-parameter ($\beta$)
gauge family interpolating between the chiral limit and the usual
conformal gauge in string theory. This idea was used to compute scattering
amplitudes of tensile chiral strings, which are given by standard
string amplitudes with modified ($\beta$-dependent) antiholomorphic
propagators.

Due to the absence of a sensible definition of the integrated vertex
operator, there is still no ordinary prescription for higher than
$3$-point amplitude computations directly from the chiral model.
The exception is the tensionless limit.

In this work this gap will be filled. Starting with a chiral string
action, the integrated vertex operator is defined, relying on the
so-called sectorized interpretation. As it turns out, this construction
effectively emulates a left/right factorization of the scattering
amplitude and introduces a relative sign flip in the propagator for
the \emph{sector-split} target space coordinates. $N$-point tree-level
amplitudes can be easily shown to coincide with the results of Siegel
\emph{et al}.

\pagebreak{}

\tableofcontents{}
\end{abstract}

\section{Introduction\label{sec:Introduction}}

The ambitwistor string was proposed in \cite{Mason:2013sva} as a
chiral string model underpinning the so-called CHY formul{\ae} \cite{Cachazo:2013hca}.
In spite of possessing an involved geometrical interpretation, the
model has a simple realization. In the bosonic case, for instance,
its chiral action takes the gauge fixed form
\begin{equation}
S_{bos}=\frac{1}{2\pi}\int d^{2}z\{P_{m}\bar{\partial}X^{m}+b\bar{\partial}c+\tilde{b}\bar{\partial}\tilde{c}\},\label{eq:actionbosfixed}
\end{equation}
where $X^{m}$ denote the target space coordinates, $c$ is the holomorphic
component of reparametrization ghost and $\tilde{c}$ is the ghost
associated to the massless constraint $H\equiv P^{m}P_{m}=0$. Incidentally,
this action resembled the one proposed earlier by Hohm, Siegel and
Zwiebach (HSZ) \cite{Hohm:2013jaa}, which was considered in the context
of doubled-coordinate field theory towards a model with manifest T-duality.
The ambitwistor string, however, has no dimensionful parameter.

The connection between the ambitwistor string and the ordinary string
was not precisely understood at first. Due to the massless constraint
and its finite spectrum, it might be more intuitive to picture the
ambitwistor string as arising from some infinite tension (zero length)
limit of the string. In this case, the string shrinks to a point and
massive states become infinitely heavy, decoupling from the theory.
In terms of the action, this idea is supported classically when the
infinite tension limit is accompanied by singular field redefinitions
\cite{Mason:2013sva,Bandos:2014lja}.

On the other hand, the infinite length limit of the string yields
too a massless spectrum. The difference is that it consists of an
infinite number of higher spin states. From this point of view, if
the ambitwistor string is supposed to come from a zero tension limit,
it has to be concomitant with some process that truncates the physical
spectrum. In \cite{Siegel:2015axg}, Siegel proposed a one-parameter
family of gauges that interpolates between the usual conformal gauge
and the chiral gauge leading to the action above. In terms of the
metric $g_{ij}$, where $\{i,j\}=\tau,\sigma$ are the standard worldsheet
directions, it corresponds to
\begin{equation}
g_{ij}=e^{2\omega}\left(\begin{array}{cc}
-1+\beta & \beta\\
\beta & 1+\beta
\end{array}\right),
\end{equation}
where $\omega$ is a conformal factor and $\beta\geq0$ is the parameter.
The conformal gauge is recovered when $\beta=0$, while the chiral
limit is obtained when $\beta\to\infty$. The action \eqref{eq:actionbosfixed}
is the outcome of the BRST quantization of the first order form of
the Polyakov action in this gauge (referred to as HSZ). At least at
the classical level, the ambitwistor string can then be seen as the
tensionless limit of the usual string in the singular HSZ gauge\footnote{Singular because, strictly speaking, the limit $\beta\to\infty$ is
not a valid gauge choice since it cannot be inverted. Any finite $\beta$,
however, can be put back in the conformal gauge through an extra coordinate
transformation.}. This idea seems to be directly connected with the results of Gamboa
\emph{et al} \cite{Gamboa:1989zc,Gamboa:1989px}, where different
ordering prescriptions lead to different spectra in the tensionless
limit of the (spinning) string, called null string. Later, in \cite{Casali:2016atr},
this connection was made much more clear and precise.

In addition to the bosonic model, the extensions of the ambitwistor
string with $\mathcal{N}=1$ (heterotic) and $\mathcal{N}=2$ (types
IIA and IIB) worldsheet supersymmetry were also analyzed in \cite{Mason:2013sva}.
Their descriptions with manifest spacetime supersymmetry appeared
soon after in \cite{Berkovits:2013xba}, using the pure spinor formalism.
If compact spacetime dimensions are analyzed, naturally introducing
a dimensionful parameter, the physical spectrum is much richer and
an infinite number of higher spin massive states are present \cite{Casali:2017mss}.

For the type II case, the ambitwistor spectrum coincides with the
usual massless superstring spectrum ($\mathcal{N}=2$, $d=10$ supergravity)
and, again, the infinite tension picture seems adequate. The same
agreement is observed for the gauge sector of the heterotic model.
On the other hand, both the bosonic model and the gravity sector of
the heterotic model do not match their usual string counterparts.
In fact, a more thorough analysis of their BRST cohomology shows that
the physical spectrum is non-unitary. In particular, the kinetic action
associated to these states involves higher derivatives. This was only
recently demonstrated \cite{Berkovits:2018jvm}. 

This asymmetry between the gravity sector of the bosonic, heterotic
and type II ambitwistor spectrum is more easily understood when the
tensile picture is considered. In this case, the massless ambitwistor
constraint is replaced by
\begin{equation}
P_{m}P^{m}+\mathcal{T}^{2}\partial X_{m}\partial X^{m}=0,\label{eq:tensileconstraint}
\end{equation}
where $\mathcal{T}$ is the string tension. In principle, this tensile
``deformation'' allows for massive states in the spectrum. However,
this depends on the amount of supersymmetry of the model. For $\mathcal{N}=2$,
the physical spectrum of the chiral string is independent of the tension.
For $\mathcal{N}=1$ (heterotic), the physical spectrum contains the
usual supergravity states plus a multiplet resembling the first massive
level of the open superstring with mass $m^{2}=4\mathcal{T}$. For
$\mathcal{N}=0$ (bosonic), in addition to the usual gravity states
(graviton, dilaton and Kalb-Ramond form), the spectrum contains two
spin $2$ fields with mass $m^{2}=\pm4\mathcal{T}$. 

As proposed in \cite{Siegel:2015axg}, the tensile chiral string can
be obtained from a singular gauge limit of the usual string together
with a change in the worldsheet boundary conditions, which effectively
amounts to a sign flip in the antiholomorphic part of the $\left\langle XX\right\rangle $
propagator. This idea was explored by Huang\emph{ et al} in \cite{Huang:2016bdd},
where a modified Kawai-Lewellen-Tye factorization was used to compute
tree level amplitudes. As expected, the poles of these amplitudes
coincide with the physical spectrum described above. In the tensionless
limit, the massive spin $2$ fields of the bosonic and the heterotic
model become massless and mix with the graviton. They can be understood
as auxiliary fields responsible for introducing the higher derivative
equations of motion for the original massless states. This construction
has been demonstrated in the bosonic model in \cite{Azevedo:2019zbn},
including current algebra extensions.

The first loop computations using the ambitwistor string appeared
in \cite{Adamo:2013tsa}, indicating the one-loop modular invariance
of the GSO projected type II scattering amplitudes. Results on two-loops
appeared more recently in \cite{Geyer:2018xwu}. It seems, however,
that the purely bosonic model does not share such feature. This was
analyzed within the framework of null string theory in \cite{Yu:2017bpw}
and the results agree with the tensile model of \cite{Lee:2017utr}.
 Unfortunately, modular invariance cannot yet be confirmed using the
pure spinor construction of \cite{Berkovits:2013xba}. The reason
is that the $b$ ghost in the pure spinor formalism is a composite
operator and its definition in the tensionless limit has not been
found so far. In the tensile model, however, the $b$ ghost was successfully
built in \cite{Jusinskas:2016qjd}, which introduced the sectorized
description of the chiral models. The idea behind this interpretation
is that a chiral model with action of the form \eqref{eq:actionbosfixed}
can be effectively described in terms of two sectors that emulate
the usual left and right movers of the (super) string.

Since the pure spinor formalism still lacked a fundamental worldsheet
description, the relation between the chiral actions of \cite{Jusinskas:2016qjd}
and \cite{Siegel:2015axg} was not very clear. This was only completely
achieved later in \cite{Jusinskas:2019vmd}, where the equivalent
of the HSZ gauge was implemented with the derivation of the chiral
action of \cite{Jusinskas:2016qjd}. Another issue that prevented
a more immediate identification of the sectorized model with Siegel's
chiral string was the absence of the integrated vertex operator (IVO)
in the former. Therefore, only $3$-point tree-level amplitudes could
be directly compared. 

In general, chiral models lack a natural ingredient to define an IVO
as an integration over a Riemann surface, since all fields are holomorphic.
In the ambitwistor string, this obstacle is overcome with the insertion
of the BRST-closed operators $\bar{\delta}(k^{m}P_{m})$, where $k^{m}$
is the linear momentum of the vertex. In turn, this operator enforces
the localization of the scattering amplitudes. The definition of the
operator $\bar{\delta}(k^{m}P_{m})$ was made more precise by Ohmori
in \cite{Ohmori:2015sha} and can be written instead as $\bar{\delta}(H_{-1})$,
where $H_{-1}$ is the mode $-1$ of $H$ acting on the unintegrated
vertex operator. It coincides with $\bar{\delta}(k^{m}P_{m})$ when
the $X^{m}$ dependence of the vertex operator is only through $e^{ik\cdot X}$.

In this work, Ohmori's construction will be generalized in order to
properly define the IVO of the tensile chiral string. The main ingredient
in this construction is the sector-splitting operator $\bar{\Delta}$,
formally defined as
\begin{equation}
\bar{\Delta}\equiv e^{i\bar{z}\mathcal{H}_{-1}},\label{eq:Dbar}
\end{equation}
where $\mathcal{H}_{-1}$ is the $-1$ mode of the BRST-exact generalization
of the left-hand side of equation \eqref{eq:tensileconstraint}. This
operator has a simple interpretation in the sectorized description,
generating a point-splitting between operators of different sectors.
In particular, its action on the target space coordinate $X^{m}$
generates a scalar field of the form $\bar{X}^{m}(z^{+},z^{-})\equiv\bar{\Delta}\cdot X^{m}(z)$,
where the coordinates $z^{\pm}=z\pm i\mathcal{T}\bar{z}$ can be interpreted
as conjugate to each other\footnote{As long as the moduli space of $z$ and $\bar{z}$ corresponds to
the complex plane, $\bar{z}$ does not necessarily have to be the
complex conjugate of $z$. This notation was chosen for the sake of
simplicity. The reader may find useful the discussion in \cite{Ohmori:2015sha}
regarding the analysis of the moduli space of chiral strings.}. It satisfies the OPE
\begin{equation}
\bar{X}^{m}(z^{+},z^{-})\,\bar{X}^{n}(0,0)\sim\tfrac{\eta^{mn}}{2\mathcal{T}}\ln\left(\tfrac{z^{-}}{z^{+}}\right),\label{eq:XbXbOPE}
\end{equation}
emulating the boundary condition change proposed by Siegel in \cite{Siegel:2015axg}.
Following this result, a series of identifications is put forward,
resulting in a simple map between the chiral models with first order
action of the form \eqref{eq:actionbosfixed} and ordinary second
order string actions in the conformal gauge but with the particular
OPE above. It is then straightforward to show that tree level scattering
amplitudes in the sectorized model are given by the ``flipped''
Kawai-Lewellen-Tye factorized amplitudes of \cite{Huang:2016bdd}.
In order to see the tensionless limit, it is enough to show that
\begin{equation}
\lim_{\mathcal{T}\to0}\bar{\Delta}\cdot X^{m}(z)=X^{m}(z)+i\bar{z}P^{m}(z),
\end{equation}
and the modular integration of $\bar{z}$ in the IVO's generates the
localization operators $\bar{\delta}(k^{m}P_{m})$.

It is worth to point out that a rigorous way of defining $N$-point
amplitudes in these chiral models already exists and it is given in
terms of intersection numbers of cohomology classes on the moduli
space of punctured Riemann surfaces (see \cite{Mizera:2019gea} and
references therein). Their equivalence with the amplitudes of \cite{Huang:2016bdd}
was already demonstrated in \cite{Mizera:2017rqa} using intersection
theory\footnote{I would like to thank Sebastian Mizera for bringing this to my attention}.
On the other hand, the definition of the integrated vertex operator
for the chiral models brings this equivalence closer to the usual
construction in string theory, with a more physical approach.

This work is organized as follows. Section \ref{sec:Review} consists
of a review of Siegel's gauge family of \cite{Siegel:2015axg} and
the sectorized interpretation in the chiral limit. Section \ref{sec:vertexop}
introduces the operator \eqref{eq:Dbar} and its properties, finally
leading to the definition of the integrated vertex operator and the
computation of $N$-point tree level scattering amplitudes directly
from the chiral model. Section \ref{sec:Final} presents some quick
final remarks and possible directions to follow. In the appendix,
the bosonic chiral string is analyzed as an example. The $4$-point
amplitude with external massless states is computed and some properties
are verified, \emph{e.g.} M\"{o}bius invariance and the existence
of physical poles only at $m^{2}=0$ and $m^{2}=\pm4\mathcal{T}$.

\section{Review of chiral strings\label{sec:Review}}

In this section, the singular gauge fixing leading to chiral strings
will be reviewed in the context of Siegel's gauge family and the sectorized
description.

\subsection{The Polyakov action and Siegel's gauge family\label{subsec:Polyakov1st}}

The Polyakov action in the Hamiltonian form is given by
\begin{multline}
S_{P}=\frac{1}{2\pi}\int d\tau d\sigma\big\{ P_{m}\partial_{\tau}X^{m}+\tfrac{1}{4\mathcal{T}}e_{+}(P_{m}+\mathcal{T}\partial_{\sigma}X_{m})(P^{m}+\mathcal{T}\partial_{\sigma}X^{m})\\
+\tfrac{1}{4\mathcal{T}}e_{-}(P_{m}-\mathcal{T}\partial_{\sigma}X_{m})(P^{m}-\mathcal{T}\partial_{\sigma}X^{m})\big\}.\label{eq:Pol1st}
\end{multline}
where $\mathcal{T}$ is the string tension and $e_{\pm}$ denote Weyl
invariant Lagrange multipliers. They can be effectively interpreted
as the gauge fields of worldsheet reparametrization symmetry, with
generators
\begin{equation}
H_{\pm}\equiv(P_{m}\pm\mathcal{T}\partial_{\sigma}X_{m})(P^{m}\pm\mathcal{T}\partial_{\sigma}X^{m}).\label{eq:WSrepgen}
\end{equation}
The associated gauge transformations, with local parameters $c_{+}$
and $c_{-}$, are given by
\begin{equation}
\begin{array}{rcl}
\delta X^{m} & = & \tfrac{1}{2}[c_{+}(P^{m}+\mathcal{T}\partial_{\sigma}X^{m})+c_{-}(P^{m}-\mathcal{T}\partial_{\sigma}X^{m})],\\
\delta P_{m} & = & \tfrac{\mathcal{T}}{2}\partial_{\sigma}[c_{+}(P^{m}+\mathcal{T}\partial_{\sigma}X^{m})-c_{-}(P^{m}-\mathcal{T}\partial_{\sigma}X^{m})],\\
\delta e_{\pm} & = & \partial_{\tau}c_{\pm}\pm(c_{\pm}\partial_{\sigma}e_{\pm}-e_{\pm}\partial_{\sigma}c_{\pm}),
\end{array}
\end{equation}
and the usual conformal gauge is obtained by setting $e_{\pm}=-1$. 

In \cite{Siegel:2015axg}, Siegel introduced a one-parameter ($\beta$)
family of gauges expressed as
\begin{equation}
e_{\pm}=\frac{\pm\beta-1}{\beta+1}.\label{eq:siegelfamily}
\end{equation}
This gauge family can be seen as an interpolation between the conformal
gauge ($\beta=0$) and the chiral gauge ($\beta\to\infty$), which
was first proposed in the doubled-coordinate field theory context
of \cite{Hohm:2013jaa}.

\begin{figure}
\fontsize{12pt}{16pt}
\def\svgwidth{\linewidth}
\centering{\resizebox{100mm}{!}{
\begingroup%
  \makeatletter%
  \providecommand\color[2][]{%
    \errmessage{(Inkscape) Color is used for the text in Inkscape, but the package 'color.sty' is not loaded}%
    \renewcommand\color[2][]{}%
  }%
  \providecommand\transparent[1]{%
    \errmessage{(Inkscape) Transparency is used (non-zero) for the text in Inkscape, but the package 'transparent.sty' is not loaded}%
    \renewcommand\transparent[1]{}%
  }%
  \providecommand\rotatebox[2]{#2}%
  \newcommand*\fsize{\dimexpr\f@size pt\relax}%
  \newcommand*\lineheight[1]{\fontsize{\fsize}{#1\fsize}\selectfont}%
  \ifx\svgwidth\undefined%
    \setlength{\unitlength}{1329bp}%
    \ifx\svgscale\undefined%
      \relax%
    \else%
      \setlength{\unitlength}{\unitlength * \real{\svgscale}}%
    \fi%
  \else%
    \setlength{\unitlength}{\svgwidth}%
  \fi%
  \global\let\svgwidth\undefined%
  \global\let\svgscale\undefined%
  \makeatother%
  \begin{picture}(1,0.64258841)%
    \lineheight{1}%
    \setlength\tabcolsep{0pt}%
    \put(0,0){\includegraphics[width=\unitlength,page=1]{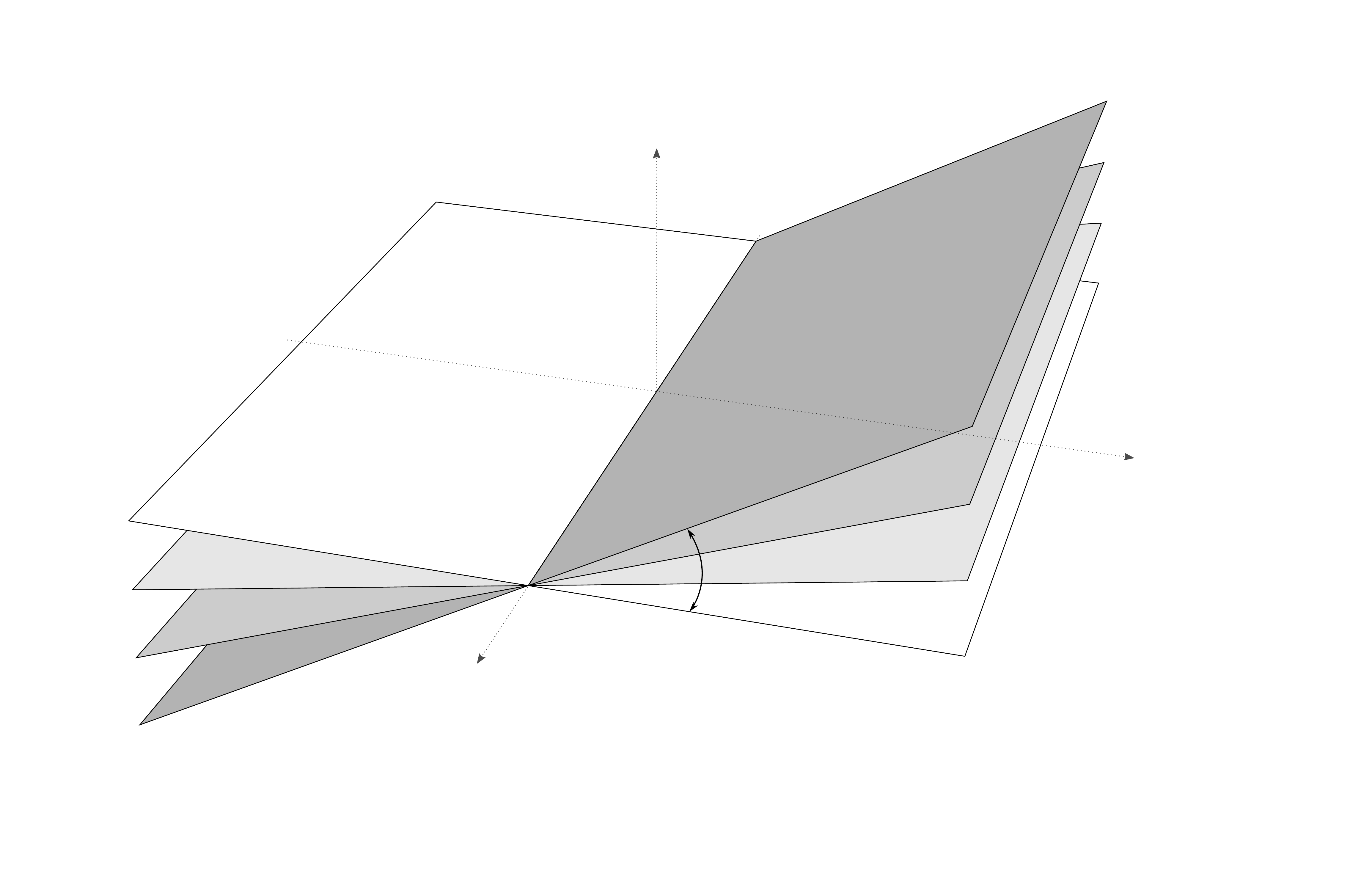}}%
    \put(0.48758464,0.5071427){\color[rgb]{0,0,0}\makebox(0,0)[lt]{\lineheight{1.25}\smash{\begin{tabular}[t]{l}$\rho$\end{tabular}}}}%
    \put(0.78897918,0.28366638){\color[rgb]{0,0,0}\makebox(0,0)[lt]{\lineheight{1.25}\smash{\begin{tabular}[t]{l}$\sigma^{+}$\end{tabular}}}}%
    \put(0.36701079,0.1662448){\color[rgb]{0,0,0}\makebox(0,0)[lt]{\lineheight{1.25}\smash{\begin{tabular}[t]{l}$\sigma^{-}$\end{tabular}}}}%
    \put(0.53971966,0.22356992){\color[rgb]{0,0,0}\rotatebox{6.02926}{\makebox(0,0)[lt]{\lineheight{1.25}\smash{\begin{tabular}[t]{l}$\theta=\arctan{\sqrt{\beta}}$\end{tabular}}}}}%
  \end{picture}%
\endgroup%
}}
\caption{Increasing $\beta$  corresponds to increasing the angle $\theta$ of the metric plane with respect to the $\sigma^+ \sigma^-$  ($\beta=0$) plane. The $\beta \to \infty$ limit  corresponds to the $\sigma^+=0$ plane.}
\label{fig:2dmetric}
\end{figure}
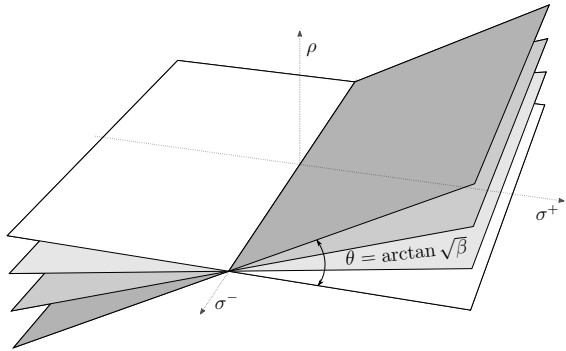

The chiral gauge is singular and has a simple geometric interpretation.
First recall that $e_{+}$ and $e_{-}$ can be expressed in terms
of the worldsheet metric $g_{ij}$ ($g^{ij}$) as
\begin{equation}
e_{\pm}\equiv\pm\frac{g^{\tau\sigma}}{g^{\tau\tau}}-\frac{1}{g^{\tau\tau}\sqrt{-g}}.\label{eq:e+-}
\end{equation}
Given the gauge choice \eqref{eq:siegelfamily}, it follows that
\begin{equation}
g_{ij}=\left(\begin{array}{cc}
-1+\beta & \beta\\
\beta & 1+\beta
\end{array}\right),\label{eq:2dmetric-beta}
\end{equation}
while $\det g_{ij}=-1$. In principle, there is also a conformal factor
but it can be conveniently tuned for a given topology using the Weyl
symmetry (for simplicity, only the flat worldsheet will be discussed
here). The line element can be written in terms of the light-cone
coordinates, $\sigma^{\pm}=\tau\pm\sigma$, as $ds^{2}=-d\sigma^{+}d\sigma^{-}+\beta(d\sigma^{+})^{2}.$
In order to picture this, it is useful to consider a three-dimensional
flat space with line element $ds^{2}=-d\sigma^{+}d\sigma^{-}+d\rho^{2}$
subjected to the constraint $\rho=\sqrt{\beta}\sigma^{+}$. The intrinsic
two-dimensional metric is then given by equation \eqref{eq:2dmetric-beta}
and corresponds to a plane inclined by an angle $\theta=\arctan\sqrt{\beta}$
with respect to $\rho=0$. See figure \ref{fig:2dmetric}.

Physically, increasing $\beta$ implies that the line element is dominated
by $d\sigma^{+}$. In the limit $\beta\to\infty$, $d\sigma^{-}$
becomes irrelevant and the worldsheet dynamics is solely governed
by the coordinate $\sigma^{+}$. For Euclidean worldsheet coordinates,
this limit leads to the (chiral) gauge fixed action
\begin{equation}
S=\frac{1}{2\pi}\int d^{2}z\{P_{m}\bar{\partial}X^{m}+b_{+}\bar{\partial}c_{+}+b_{-}\bar{\partial}c_{-}\},\label{eq:secbosaction}
\end{equation}
where the gauge parameters $c_{\pm}$ were promoted to anticommuting
ghosts with antighosts $b_{\pm}$. Note, in addition, that the string
tension $\mathcal{T}$ does not explicitly appear.

\subsection{The sectorized interpretation}

The action \eqref{eq:secbosaction} can be effectively interpreted
in terms of two sectors ``$+$'' and ``$-$'', analogous to the
usual left and right-movers of the bosonic string. This will become
clearer in the next section. To each sector, a \emph{characteristic}
energy-momentum tensor can be assigned, given by\begin{subequations}
\begin{eqnarray}
T_{+} & = & -\tfrac{1}{4\mathcal{T}}P_{m}^{+}P_{n}^{+}\eta^{mn}-2b_{+}\partial c_{+}+c_{+}\partial b_{+},\\
T_{-} & = & \tfrac{1}{4\mathcal{T}}P_{m}^{-}P_{n}^{-}\eta^{mn}-2b_{-}\partial c_{-}+c_{-}\partial b_{-},
\end{eqnarray}
\end{subequations}where
\begin{equation}
P_{m}^{\pm}\equiv P_{m}\pm\mathcal{T}\partial X_{m},\label{eq:P+-}
\end{equation}
and $\eta^{mn}$ is the flat target-space metric. The BRST charge
$Q$ manifestly expresses the sectorization as $Q=Q^{+}+Q^{-}$, such
that
\begin{equation}
Q^{\pm}\equiv\oint\{c_{\pm}T_{\pm}-b_{\pm}c_{\pm}\partial c_{\pm}\}.\label{eq:secbosBRST}
\end{equation}
As in the usual bosonic string, the number of spacetime dimensions
is $d=26$, fixed by requiring the nilpotency of $Q$.

While the combination $(T_{+}+T_{-})$ gives the left-moving Virasoro
generator, $(T_{+}-T_{-})$ takes the form of a generalized particle-like
Hamiltonian:\begin{subequations}
\begin{eqnarray}
T & = & T_{+}+T_{-},\nonumber \\
 & = & -P_{m}\partial X^{m}-b\partial c-\partial(bc)-\tilde{b}\partial\tilde{c}-\partial(\tilde{b}\tilde{c}),\\
\mathcal{H} & \equiv & \mathcal{T}(T_{+}-T_{-}),\nonumber \\
 & = & -\tfrac{1}{2}P_{m}P^{m}-\tfrac{\mathcal{T}^{2}}{2}\partial X_{m}\partial X^{m}-\mathcal{T}(b_{+}\partial c_{+}-b_{-}\partial c_{-})+\mathcal{T}\partial(c_{+}b_{+}-c_{-}b_{-}).
\end{eqnarray}
\end{subequations}Since $\{Q,b_{\pm}\}=T_{\pm}$, both $T$ and $\mathcal{H}$
are BRST-exact. Note that $T_{+}$ and $T_{-}$ may have the interpretation
of energy-momentum tensors only within their respective sector.

The BRST charge can be cast in a more traditional form as
\begin{equation}
Q=\oint\{cT-bc\partial c+\tfrac{1}{2}\tilde{c}\mathcal{H}+\ldots\},\label{eq:BRSTambitwistor}
\end{equation}
where the $\ldots$ denotes terms of order $\mathcal{T}^{1}$ and
$\mathcal{T}^{2}$, and
\begin{equation}
\begin{array}{rclcrcl}
c & \equiv & \tfrac{1}{2}(c_{+}+c_{-}), &  & \tilde{c} & \equiv & \tfrac{1}{2\mathcal{T}}(c_{-}-c_{+}),\\
b & \equiv & (b_{+}+b_{-}), &  & \tilde{b} & \equiv & \mathcal{T}(b_{-}-b_{+}).
\end{array}\label{eq:bcghosts-redefinition}
\end{equation}
$Q$ resembles the bosonic ambitwistor BRST charge proposed by Mason
and Skinner \cite{Mason:2013sva} and, in fact, can be thought of
as its tensile generalization.

The sectorized description can be extended fairly easily as long as
the chiral action of the models under consideration have vanishing
\emph{characteristic} central charges, corresponding to the absence
of quartic poles in the OPE's
\begin{equation}
T_{\pm}(z)\,T_{\pm}(y)\sim\frac{2T_{\pm}}{(z-y)^{2}}+\frac{\partial T_{\pm}}{(z-y)},
\end{equation}
where $T_{+}$ and $T_{-}$ can be generically written as
\begin{equation}
T_{\pm}=\mp\tfrac{1}{4\mathcal{T}}P_{m}^{\pm}P_{n}^{\pm}\eta^{mn}+T_{\pm}^{ghost}+T_{\pm}^{matter}.
\end{equation}
In other words, the central charge of the chiral model has to vanish
and $\mathcal{H}$ has to be a primary operator.

Some examples of interest include the spinning string with $\mathcal{N}=(1,1)$
, $\mathcal{N}=(1,0)$ and $\mathcal{N}=(0,1)$ worldsheet supersymmetry,
the pure spinor superstring with $\mathcal{N}=2$ and $\mathcal{N}=1$
spacetime supersymmetry\footnote{In the pure spinor formalism, $b_{+}$ and/or $b_{-}$ are composite
operators satisfying $\{Q,b_{\pm}\}=T_{\pm}$.} or purely bosonic models in $d<26$ extended with current algebras.
All these models contain graviton excitations (either type II, heterotic
or bosonic), with or without ghosts, and recover the corresponding
ambitwistor string in their tensionless limit.

\section{Vertex operators and amplitudes\label{sec:vertexop}}

In the chiral strings mentioned above, physical states are defined
to be in their ghost number $2$ BRST cohomology and annihilated by
$b_{0}$, the zero mode of the $b$ ghost \emph{cf}. equation \eqref{eq:bcghosts-redefinition}.
This last requirement is the chiral analogous of the condition $(b_{0}-\bar{b}_{0})=0$
for off-shell closed string states and was shown to provide the correct
prescription for ambitwistor strings \cite{Berkovits:2018jvm} (see
\cite{Reid-Edwards:2015stz,Reid-Edwards:2017goq} for more details
on the definition of off-shell states and the ambitwistor string field
theory).

The unintegrated vertex operators are straightforward to build (normal
ordering is assumed whenever needed). For momentum eigenstates, their
different components have the generic form
\begin{equation}
U(z;k^{m})=U_{+}(z)U_{-}(z)e^{ik\cdot X(z)},\label{eq:uvo}
\end{equation}
where $U_{+}$ and $U_{-}$ are composite operators constructed exclusively
from the ``$+$'' and ``$-$'' sectors respectively. As long as
the physical state conditions are satisfied, $U_{+}$ and $U_{-}$
do not necessarily have the same ghost number.

The definition of integrated vertex operators is slightly more intricate.
Because the model is chiral, the usual recipe from string theory does
not work. Naively, the integrated vertex operator could be defined
as
\begin{equation}
V(z;k^{m})\overset{?}{\equiv}(b_{-})_{-1}\cdot(b_{+})_{-1}\cdot U(z;k^{m}),
\end{equation}
where the subscript $-1$ denotes the corresponding mode of the $b$
ghost operators. Note, however, that $V$ has conformal weight $(2,0)$
and satisfies $\bar{\partial}V=0$, therefore the integration over
the Riemann surface $\Sigma$ is not well defined. In addition,
\begin{eqnarray}
[Q,V(z;k^{m})] & = & (T_{-})_{-1}\cdot(b_{+})_{-1}\cdot U(z;k^{m})-(b_{-})_{-1}\cdot(T_{+})_{-1}\cdot U(z;k^{m})\nonumber \\
 & = & \tfrac{1}{2}T_{-1}\cdot(b_{-}-b_{+})_{-1}\cdot U(z;k^{m})+\tfrac{1}{2\mathcal{T}}\mathcal{H}_{-1}\cdot(b_{+}+b_{-})_{-1}\cdot U(z;k^{m})\nonumber \\
 & = & \tfrac{1}{2}\tfrac{\partial}{\partial z}[(b_{-}-b_{+})_{-1}\cdot U(z;k^{m})]+\tfrac{1}{2\mathcal{T}}\mathcal{H}_{-1}\cdot(b_{+}+b_{-})_{-1}\cdot U(z;k^{m}).
\end{eqnarray}
While the first term on the right hand side of the last equation can
be disregarded as a boundary contribution in the \emph{moduli} space
integration, the second term does not vanish and the integrated vertex
operator is not BRST invariant.

A natural way of fixing these issues would be to formally introduce
a BRST closed operator $\bar{\delta}(\mathcal{H}_{-1})$ with conformal
weight $(-1,1)$, such that the product $\bar{\delta}(\mathcal{H}_{-1})\cdot\mathcal{H}_{-1}$
vanishes under the appropriate conditions, \emph{e.g.} corresponding
to a boundary of the \emph{moduli} space. In this case, the integrated
vertex operator should be defined as
\begin{equation}
V(z;k^{m})\equiv(b_{-})_{-1}\cdot(b_{+})_{-1}\cdot\bar{\delta}(\mathcal{H}_{-1})\cdot U(z;k^{m}),\label{eq:intvertex-prop}
\end{equation}
which satisfies $[Q,V(z;k^{m})]=\tfrac{\partial}{\partial z}(\ldots)$.
In the tensionless limit, this idea was proposed by Ohmori in \cite{Ohmori:2015sha}
and agrees with the operator $\bar{\delta}(k\cdot P)$ introduced
by Mason and Skinner in \cite{Mason:2013sva} as long as the only
dependence of the vertex operator on the target-space coordinates
is through $e^{ik\cdot X}$. The \emph{moduli} space integration is
then well defined and the role of $\bar{\delta}(k\cdot P)$ is to
impose the so-called scattering equations in the $N$-point amplitude
computations.

In this section, the integrated vertex operator for the sectorized
model will be finally built, motivated by the definition \eqref{eq:intvertex-prop}.
Besides providing a more robust and intuitive definition of the operator
$\bar{\delta}(\mathcal{H}_{-1})$, the tools developed here explain
the expected connection between the sectorized model and the $\beta\to\infty$
limit of Siegel's string \cite{Siegel:2015axg}.

\subsection{The sector-splitting operator}

Consider the BRST-closed operator
\begin{eqnarray}
\bar{\Delta} & \equiv & \sum_{n=0}^{\infty}\tfrac{1}{n!}\left(i\bar{z}\mathcal{H}_{-1}\right)^{n},\\
 & = & e^{i\bar{z}\mathcal{H}_{-1}},\label{eq:deltabar}
\end{eqnarray}
where
\begin{equation}
\mathcal{H}_{-1}=\mathcal{T}\oint\left\{ T_{+}-T_{-}\right\} .
\end{equation}

By construction,\begin{subequations}\label{eq:Deltabarproperties}
\begin{eqnarray}
[L_{0},\bar{\Delta}] & = & \bar{z}\bar{\partial}\bar{\Delta},\\
\bar{\Delta}\cdot\mathcal{H}_{-1} & = & -i\bar{\partial}\bar{\Delta},
\end{eqnarray}
\end{subequations}where $L_{0}$ is the Virasoro zero mode.

The action of the operator $\bar{\Delta}$ can be understood in terms
of the sectorized interpretation as follows. While $T=T_{+}+T_{-}$
generates holomorphic worldsheet translations,
\begin{equation}
e^{zL_{-1}}\mathcal{O}(0)=\mathcal{O}(z),
\end{equation}
there is no analogous operation for the antiholomorphic sector because
the classical equations of motion imply $\bar{\partial}=0$. As discussed
in subsection \ref{subsec:Polyakov1st}, this is a consequence of
the singular metric gauge (HSZ). The antiholomorphic dependence is
then artificially reintroduced by the operator $\bar{\Delta}$. This
operation depends on the string tension $\mathcal{T}$ and on the
sector on which the operator acts. Observe, for example, that
\begin{equation}
\bar{\Delta}\cdot P_{m}^{\pm}(z)=P_{m}^{\pm}(z\pm i\mathcal{T}\bar{z}),
\end{equation}
and, more generally,
\begin{equation}
\bar{\Delta}\cdot\mathcal{O}^{\pm}(z)=\mathcal{O}^{\pm}(z\pm i\mathcal{T}\bar{z}),
\end{equation}
where $\mathcal{O}^{+}$ ($\mathcal{O}^{-}$) denotes operators exclusively
built from the plus (minus) sector. Therefore, $\bar{\Delta}$ creates
a point-splitting between the two sectors. The presence of the string
tension in the arguments is unusual but makes it easier to determine
the tensionless limit. It could have been removed, of course, by a
redefinition of \eqref{eq:deltabar} as $\bar{\Delta}=\exp\{\tfrac{i}{\mathcal{T}}\bar{z}\mathcal{H}_{-1}\}$.
It will be convenient to define also the sector coordinates $z^{\pm}$
as
\begin{equation}
\begin{array}{cc}
z^{\pm}\equiv z\pm i\mathcal{T}\bar{z}, & \partial_{\pm}\equiv\tfrac{1}{2\mathcal{T}}(\mathcal{T}\partial\mp i\bar{\partial}),\end{array}\label{eq:z+-}
\end{equation}
such that $\partial_{\pm}z^{\pm}=1$ and $\partial_{\pm}z^{\mp}=0$.
These coordinates play the role of the usual holomorphic and anti-holomorphic
pair, but they are not suitable for the tensionless limit analysis
since they become degenerate, \emph{i.e.} $z^{\pm}\to z$ for $\mathcal{T}\to0$.

The only non-trivial operation is the action of $\bar{\Delta}$ on
the target-space coordinates, $X^{m}$, since they do not belong to
a specific sector. It will be defined as
\begin{equation}
\bar{X}_{m}(z,\bar{z})\equiv\bar{\Delta}\cdot X_{m}(z),
\end{equation}
and it is straightforward to show from the series expansion \eqref{eq:deltabar}
that
\begin{eqnarray}
\bar{X}_{m}(z,\bar{z}) & = & X_{m}(z)+\tfrac{1}{2}\sum_{n=1}^{\infty}\tfrac{i\bar{z}}{n!}\left(i\bar{z}\mathcal{T}\right)^{n-1}\left\{ \partial^{n-1}P_{m}^{+}(z)+(-1)^{n-1}\partial^{n-1}P_{m}^{-}(z)\right\} ,\nonumber \\
 & = & \tfrac{1}{2}X_{m}(z+i\mathcal{T}\bar{z})+\tfrac{1}{2}X_{m}(z-i\mathcal{T}\bar{z})+i\bar{z}\sum_{n=0}^{\infty}\tfrac{\left(i\bar{z}\mathcal{T}\right)^{2n}}{(2n+1)!}\partial^{2n}P_{m}(z).\label{eq:Xbarseries}
\end{eqnarray}
To obtain the second line from the first line, it suffices to recall
the definition \eqref{eq:P+-}. Note that $\bar{X}_{m}(z,\bar{z})$
satisfies the equations of motion\begin{subequations}
\begin{eqnarray}
\bar{\partial}\bar{X}_{m}(z,\bar{z}) & = & \tfrac{1}{2}P_{m}^{+}(z+i\mathcal{T}\bar{z})+\tfrac{1}{2}P_{m}^{-}(z-i\mathcal{T}\bar{z}),\\
\partial\bar{X}_{m}(z,\bar{z}) & = & \tfrac{1}{2\mathcal{T}}P_{m}^{+}(z+i\mathcal{T}\bar{z})-\tfrac{1}{2\mathcal{T}}P_{m}^{-}(z-i\mathcal{T}\bar{z}),
\end{eqnarray}
\end{subequations}which can be rewritten in a more suggestive form
in terms of the coordinates $z^{\pm}$,\begin{subequations}\label{eq:ddbarX}
\begin{eqnarray}
\partial_{+}\bar{X}_{m} & = & \tfrac{1}{2\mathcal{T}}P_{m}^{+}(z^{+}),\\
\partial_{-}\bar{X}_{m} & = & -\tfrac{1}{2\mathcal{T}}P_{m}^{-}(z^{-}),\\
\partial_{+}\partial_{-}\bar{X}_{m} & = & 0.
\end{eqnarray}
\end{subequations}The BRST transformation of $\bar{X}^{m}$ also
incorporates such sector splitting as
\begin{eqnarray}
[Q,\bar{X}_{m}(z,\bar{z})] & = & \tfrac{1}{2\mathcal{T}}c_{+}P_{m}^{+}(z+\bar{z}\mathcal{T})-\tfrac{1}{2\mathcal{T}}c_{-}P_{m}^{-}(z-\bar{z}\mathcal{T}),\nonumber \\
 & = & c_{+}\partial_{+}\bar{X}_{m}(z^{+})+c_{-}\partial_{-}\bar{X}_{m}(z^{-}).
\end{eqnarray}

All these results suggest that $\bar{X}^{m}$ behaves like an ordinary
worldsheet scalar from a second order action, including a non-trivial
OPE with itself which can be computed to be
\begin{equation}
\bar{X}^{m}(z^{+},z^{-})\,\bar{X}^{n}(y^{+},y^{-})\sim-\tfrac{\eta^{mn}}{2\mathcal{T}}\ln(z^{+}-y^{+})+\tfrac{\eta^{mn}}{2\mathcal{T}}\ln(z^{-}-y^{-}),\label{eq:Siegelprop}
\end{equation}
The catch is that the would-be holomorphic and antiholomorphic sectors
have a propagator with opposite sign, similar to Siegel's proposal
of \cite{Siegel:2015axg}. The demonstration of \eqref{eq:Siegelprop}
is straightforward. From the second line of \eqref{eq:Xbarseries},
it is easy to see that the singular terms contributing to the OPE
can be organized as
\begin{multline}
\bar{X}^{m}(z^{+},z^{-})\,\bar{X}^{n}(y^{+},y^{-})\sim\frac{1}{4\mathcal{T}}\sum_{k=1}^{\infty}\tfrac{1}{k!}\left(i\bar{y}\mathcal{T}\right)^{k}\left\{ \left[X^{m}(z+i\mathcal{T}\bar{z})+X^{m}(z-i\mathcal{T}\bar{z})\right]\left[\partial^{k-1}P^{n}(y)\right]\right\} \\
-\frac{1}{4\mathcal{T}}\sum_{k=1}^{\infty}\tfrac{1}{k!}\left(-i\bar{y}\mathcal{T}\right)^{k}\left\{ \left[X^{m}(z+i\mathcal{T}\bar{z})+X^{m}(z-i\mathcal{T}\bar{z})\right]\left[\partial^{k-1}P^{n}(y)\right]\right\} \\
+\frac{1}{4\mathcal{T}}\sum_{k=1}^{\infty}\tfrac{1}{k!}\left(i\bar{z}\mathcal{T}\right)^{k}\left\{ \left[\partial^{k-1}P^{m}(z)\right]\left[X^{n}(y+i\mathcal{T}\bar{y})+X^{n}(y-i\mathcal{T}\bar{y})\right]\right\} \\
-\frac{1}{4\mathcal{T}}\sum_{k=1}^{\infty}\tfrac{1}{k!}\left(-i\bar{z}\mathcal{T}\right)^{k}\left\{ \left[\partial^{k-1}P^{m}(z)\right]\left[X^{n}(y+i\mathcal{T}\bar{y})+X^{n}(y-i\mathcal{T}\bar{y})\right]\right\} 
\end{multline}
Next, using the OPE
\begin{equation}
X^{m}(z)\,P_{n}^{\pm}(y)\sim\frac{\delta_{n}^{m}}{(z-y)},
\end{equation}
the equation above can be rewritten as
\begin{multline}
\bar{X}^{m}(z^{+},z^{-})\,\bar{X}^{n}(y^{+},y^{-})\sim\frac{\eta^{mn}}{4\mathcal{T}}\sum_{k=1}^{\infty}\frac{1}{k}\left(i\bar{y}\mathcal{T}\right)^{k}\left\{ \frac{1}{(z+i\mathcal{T}\bar{z}-y)^{k}}+\frac{1}{(z-i\mathcal{T}\bar{z}-y)^{k}}\right\} \\
-\frac{\eta^{mn}}{4\mathcal{T}}\sum_{k=1}^{\infty}\frac{1}{k}\left(-i\bar{y}\mathcal{T}\right)^{k}\left\{ \frac{1}{(z+i\mathcal{T}\bar{z}-y)^{k}}+\frac{1}{(z-i\mathcal{T}\bar{z}-y)^{k}}\right\} \\
+\frac{\eta^{mn}}{4\mathcal{T}}\sum_{k=1}^{\infty}\frac{1}{k}\left(i\bar{z}\mathcal{T}\right)^{k}\left\{ \frac{1}{(y+i\mathcal{T}\bar{y}-z)^{k}}+\frac{1}{(y-i\mathcal{T}\bar{y}-z)^{k}}\right\} \\
-\frac{\eta^{mn}}{4\mathcal{T}}\sum_{k=1}^{\infty}\frac{1}{k}\left(-i\bar{z}\mathcal{T}\right)^{k}\left\{ \frac{1}{(y+i\mathcal{T}\bar{y}-z)^{k}}+\frac{1}{(y-i\mathcal{T}\bar{y}-z)^{k}}\right\} 
\end{multline}
These series correspond to the Taylor expansion of logatithm functions.
For example,
\begin{equation}
\sum_{k=1}^{\infty}\frac{1}{k}\frac{\left(i\bar{y}\mathcal{T}\right)^{k}}{(z+i\mathcal{T}\bar{z}-y)^{k}}=\ln\left(z+i\mathcal{T}\bar{z}-y+i\mathcal{T}\bar{y}\right)-\ln\left(z+i\mathcal{T}\bar{z}-y\right).
\end{equation}
Therefore, after some trivial cancellations, \eqref{eq:Siegelprop}
is obtained. Reproducing such structure in a chiral model ($\beta\to\infty$
limit in Siegel's gauge family) was only possible due to the introduction
of the sector-splitting operator.

\subsection{Integrated vertex operator and tree level amplitudes\label{subsec:ivoscattering}}

Having introduced the sector-splitting operator, the next step is
to analyze its action on the unintegrated vertex operators of the
form \eqref{eq:uvo}:
\begin{eqnarray}
\bar{U}(z^{+},z^{-};k) & \equiv & \bar{\Delta}\cdot U(z;k)\nonumber \\
 & = & U_{+}(z^{+})U_{-}(z^{-})e^{ik\cdot\bar{X}(z^{+},z^{-})}.\label{eq:uvobar}
\end{eqnarray}
As a consequence,
\begin{equation}
U(z;k)=\lim_{z^{\pm}\to z}\bar{U}(z^{+},z^{-};k).
\end{equation}
Using the equations \eqref{eq:ddbarX}, it is then trivial to map
any vertex operator in (the chiral limit of) Siegel's string to the
sectorized model and vice-versa.

The integrated vertex operator in the sectorized model will be defined
as
\begin{equation}
V(z,\bar{z})=(b_{-})_{-1}\cdot(b_{+})_{-1}\cdot\bar{\Delta}\cdot U,\label{eq:ivo}
\end{equation}
and satisfies\begin{subequations}
\begin{eqnarray}
[L_{0},V] & = & V+\bar{\partial}(\bar{z}V),\\{}
[\bar{L}_{0},V] & = & V-\bar{\partial}(\bar{z}V),\\{}
[Q,V] & = & \tfrac{1}{2}\{Q,(b_{+}+b_{-})_{-1}\}\cdot(b_{+}-b_{-})_{-1}\cdot\bar{U}\nonumber \\
 &  & -\tfrac{1}{2}(b_{+}+b_{-})_{-1}\cdot\{Q,(b_{+}-b_{-})_{-1}\}\cdot\bar{U}\nonumber \\
 & = & \partial\{\tfrac{1}{2}(b_{+}-b_{-})_{-1}\cdot\bar{U}\}+i\bar{\partial}\{\tfrac{1}{2}(b_{+}+b_{-})_{-1}\cdot\bar{U}\},
\end{eqnarray}
\end{subequations}where $\bar{L}_{0}=-\bar{z}\bar{\partial}$ and
the $\bar{\Delta}$ properties \eqref{eq:Deltabarproperties} were
used. Inside the \emph{moduli} space integration, $V$ has the expected
conformal transformation and its BRST transformation can be written
in terms of boundary contributions that vanish upon integration over
compact Riemann surfaces.

Naturally, the tensionless limit of this operator should agree with
\cite{Mason:2013sva,Ohmori:2015sha}. In particular, it would be interesting
to reproduce Mason and Skinner's $\bar{\delta}(k\cdot P)$ insertions
which impose the scattering equations. Through equation \eqref{eq:Xbarseries},
it follows that the tensionless limit of $\bar{X}_{m}$ is given by
$X_{m}(z)+i\bar{z}P_{m}(z)$. Consequently,
\begin{equation}
\lim_{\mathcal{T}\to0}\bar{\Delta}\cdot e^{ik\cdot X}=:e^{ik\cdot X}e^{-\bar{z}(k\cdot P)}:.
\end{equation}
 When integrated over $\bar{z}$, the second exponential can be thought
of as a representation for $\bar{\delta}(k\cdot P)$ and the integrated
vertex \eqref{eq:ivo} has the expected ambitwistor string limit.

At tree level, $N$-point amplitudes can be cast as
\begin{equation}
\mathcal{A}_{N}(k^{1},\ldots,k^{N})=\left\langle \prod_{i=1}^{3}U_{i}(z_{i};k^{i})\prod_{j=4}^{N}\mathcal{V}_{j}(k^{j})\right\rangle ,\label{eq:Namplitude}
\end{equation}
where
\begin{eqnarray}
\mathcal{V}_{j}(k^{j}) & \equiv & \int_{S^{2}}d^{2}z_{j}\,V(z_{j},\bar{z}_{j};k^{j}),\nonumber \\
 & = & \tfrac{1}{2\mathcal{T}}\int_{S^{2}}dz_{j}^{+}dz_{j}^{-}V'(z_{j}^{+},z_{j}^{-};k^{j}),\label{eq:ivowithintegral}
\end{eqnarray}
and $V'=V$ when $z_{j}^{\pm}=z_{j}\pm i\mathcal{T}\bar{z}_{j}$.
Because of the point-splitting generated by the operator $\bar{\Delta}$,
it might be more convenient to express $\mathcal{V}_{j}$ in terms
of the integration over $z_{j}^{\pm}$, but this is valid only for
$\mathcal{T}\neq0$.

The actual computation of $\mathcal{A}_{N}$ is very similar to the
one in ordinary string theory and, in fact, \emph{identical} to the
amplitudes of Huang, Siegel and Yuan in \cite{Huang:2016bdd}. The
demonstration goes as follows. Consider the unintegrated vertex operators
\eqref{eq:uvobar} and
\begin{equation}
\mathcal{\bar{A}}_{N}(k^{1},\ldots,k^{N})\equiv\left\langle \prod_{i=1}^{3}\bar{U}_{i}(z_{i}^{+},z_{i}^{-};k^{i})\prod_{j=4}^{N}\mathcal{V}_{j}(k^{j})\right\rangle .\label{eq:Nbaramplitude}
\end{equation}
In this amplitude, the plus and minus sectors behave like the usual
holomorphic and antiholomorphic sectors of the string. Using equations
\eqref{eq:ddbarX}, any vertex in the sectorized description can be
mapped to Siegel's string in the chiral limit. From the conformal
field theory point of view, evaluating $\mathcal{\bar{A}}_{N}(k^{1},\ldots,k^{n})$
using the $(P_{m},X^{m})$ first order action is equivalent to the
same computation using the second order action for $\bar{X}^{m}$
with propagator \eqref{eq:Siegelprop}. Since $\mathcal{\bar{A}}_{N}$
is M\"{o}bius invariant, \emph{i.e.} independent of the fixed coordinates
$z_{i}^{\pm}$ for $i=1,2,3$, setting $z_{i}^{+}=z_{i}^{-}$ would
then recover the amplitude $\mathcal{A}_{N}$. The only difference
is that in the chiral model there are no $\beta$-modified propagators.

It might be useful to analyze the $(P_{m},X^{m})$ contribution to
$\mathcal{A}_{N}(k^{1},\ldots,k^{N})$. Through the equations \eqref{eq:ddbarX}
and the OPE \eqref{eq:Siegelprop}, it is easy to show that
\begin{equation}
\bar{X}_{m}(z^{+},z^{-})\,P_{n}^{\pm}(y)\sim\frac{\eta_{mn}}{(z^{\pm}-y)},\label{eq:XbarP+-OPE}
\end{equation}
which can be used to compute the OPE reduction of $\mathcal{A}_{N}(k^{1},\ldots,k^{N})$
by contracting all the operators $P_{m}^{\pm}$ with themselves and
with the exponentials. Finally, the remaining contribution involves
only plane-wave operators and can be generically written as
\begin{equation}
\mathcal{A}_{N}^{X}(z,\bar{z},k)=\left\langle \prod_{i=1}^{3}:e^{ik^{i}\cdot X(z_{i})}:\prod_{j=4}^{N}:e^{ik^{j}\cdot\bar{X}(z_{j},\bar{z}_{j})}:\right\rangle .
\end{equation}

There are two simple ways for evaluating this contribution. The first
is to analyze instead
\begin{equation}
\mathcal{\bar{A}}_{N}^{X}(z^{+},z^{-},k)=\left\langle \prod_{i=1}^{N}:e^{ik^{i}\cdot\bar{X}(z_{i}^{+},z_{i}^{-})}:\right\rangle ,
\end{equation}
using the propagator \eqref{eq:Siegelprop}. The zero-mode integration
of $X^{m}$ gives the usual momentum conservation delta,
\begin{equation}
\delta^{d}\left(\sum_{i=1}^{N}k_{i}^{m}\right),\label{eq:deltaconservation}
\end{equation}
and the final result can be sketched as
\begin{equation}
\mathcal{\bar{A}}_{N}^{X}(z^{+},z^{-},k)\propto\delta^{d}\left(\sum k\right)\prod_{i>j}^{N}\left(\frac{z_{ij}^{+}}{z_{ij}^{-}}\right)^{\frac{(k_{i}\cdot k_{j})}{2\mathcal{T}}}.\label{eq:AbarXPamplitude}
\end{equation}
As explained above, $\mathcal{A}_{N}^{X}$ is recovered for $z_{i}^{+}=z_{i}^{-}$
and $i=1,2,3$.

Alternatively, $\mathcal{A}_{N}^{X}$ can be computed in a similar
way to the standard ambitwistor construction. By incorporating the
exponentials in the path integral action, it can be cast as
\begin{multline}
S[X,P]=\int d^{2}z\Bigg\{\tfrac{1}{2\pi}P_{m}\bar{\partial}X^{m}+i\sum_{i=1}^{3}k_{m}^{i}X^{m}\delta^{2}(z-z_{i})+\tfrac{i}{2}\sum_{j=4}^{N}k_{m}^{j}X^{m}[\delta^{2}(z-z_{j}^{+})+\delta^{2}(z-z_{j}^{-})]\\
-\bar{z}\sum_{j=4}^{N}\sum_{n=0}^{\infty}\tfrac{\left(i\bar{z}\mathcal{T}\right)^{2n}}{(2n+1)!}k^{jm}\partial^{2n}P_{m}(z)\,\delta^{2}(z-z_{j})\Bigg\},\label{eq:actionXP}
\end{multline}
where equation \eqref{eq:Xbarseries} was used. Again, the zero-mode
integration of $X^{m}$ gives \eqref{eq:deltaconservation}, while
the integration of the non-zero modes imply that
\begin{equation}
\tfrac{1}{2\pi}\bar{\partial}P_{m}=i\sum_{i=1}^{3}k_{m}^{i}\delta^{2}(z-z_{i})+\tfrac{i}{2}\sum_{j=4}^{N}k_{m}^{j}[\delta^{2}(z-z_{j}-i\mathcal{T}\bar{z}_{j})+\delta^{2}(z-z_{j}+i\mathcal{T}\bar{z}_{j})].
\end{equation}
On the Riemann sphere, there is a unique solution for this equation,
given by
\begin{equation}
P^{m}(z)=i\sum_{i=1}^{3}\frac{k_{i}^{m}}{(z-z_{i})}+i\sum_{j=4}^{N}\left(\frac{\tfrac{1}{2}k_{j}^{m}}{(z-z_{j}-i\mathcal{T}\bar{z}_{j})}+\frac{\tfrac{1}{2}k_{j}^{m}}{(z-z_{j}+i\mathcal{T}\bar{z}_{j})}\right),\label{eq:Pmsolution}
\end{equation}
which has the expected tensionless limit. Because of the point-splitting,
$P_{m}(z_{j})$ is not singular for $\mathcal{T}\neq0$. It is then
straightforward to show that the amplitude contribution of the action
$S[X,P]$ is equal to \eqref{eq:AbarXPamplitude} when $z_{i}^{+}=z_{i}^{-}$.
In particular, the Koba-Nielsen like factors of the amplitude are
generated by replacing the solution \eqref{eq:Pmsolution} on the
second line of \eqref{eq:actionXP}.

\section{Final remarks\label{sec:Final}}

The results presented here establish a simple realization of the fact
that tensile chiral strings can be effectively seen as ordinary strings
in the conformal gauge with a relative sign flip between the left
and right moving parts of the $\left\langle XX\right\rangle $ propagator.
This idea was proposed by Siegel in \cite{Siegel:2015axg} and interpreted
as a change in the boundary conditions of the model.

The sectorized description of \cite{Jusinskas:2016qjd}, therefore,
was an incomplete manifestation of this equivalence. The missing link,
presented here in section \ref{sec:vertexop}, was the explicit construction
of the integrated vertex operators in the chiral model.

At genus $0$, $N$-point amplitudes computed using either the new
definition of the integrated vertex operator or the Siegel's sign-flipped
model are equivalent. At the loop level, on the other hand, there
might be some subtleties. The modular invariance of the type II ambitwistor
string at one-loop was shown in \cite{Adamo:2013tsa}. However, for
the bosonic case in the null string framework, one-loop modular invariance
was not observed \cite{Yu:2017bpw}. Using the sign-flipped model,
this result was confirmed for tensile chiral strings \cite{Lee:2017utr}.
It might be interesting to investigate these results within the framework
presented here. In particular, a better understanding of the geometrical
meaning of Siegel's gauge family for the torus and higher genus Riemann
surfaces should be useful. It is not obvious, for example, that the
mentioned equivalence between the tensile chiral string and the ordinary
string with the sign flipped propagator holds at loop level. The reason
is that the point-splitting generated by the operator $\bar{\Delta}$
is invisible to the modular parameter. In other words, the would-be
left and right moving propagators depend on the same modular parameter,
$\tau$, as opposite to the usual $\tau$ and $\bar{\tau}$. An appropriate
loop prescription should be found and, in this direction, the results
of \cite{Casali:2017zkz} may be helpful.

The solutions \eqref{eq:Pmsolution} give rise to modified scattering
equations that should also accommodate massive particles. The momenta
$k_{j}$ ($j\geq4$) have a split structure, each one virtually behaving
as two particles with half of the total momentum located on the Riemann
sphere at $z_{j}\pm i\mathcal{T}\bar{z}_{j}$. However, this idea
has not yet been investigated and will be left to a future work. The
possible outcomes shall be compared with known results in the literature
of scattering equations of massive particles, \emph{e.g.} \cite{Naculich:2014naa}.

\textbf{Acknowledgments:} I would like to thank Matheus Lize, and especially
Thales Azevedo and the anonymous referee from JHEP for useful comments and suggestions. I would like to
thank also the Galileo Galilei Institute for Theoretical Physics and
INFN for their hospitality and support during the workshop \textquotedbl String
Theory from a worldsheet perspective\textquotedbl , where part of
this work has been done. This research has been supported by the Czech
Science Foundation - GA\v{C}R, project 19-06342Y.

\appendix

\section{A concrete example: the bosonic chiral string\label{sec:Example}}

In this appendix, the ideas developed in the main text will be applied
to the bosonic chiral string, with BRST charge \eqref{eq:secbosBRST}.
The cohomology will be reviewed with a subsequent evaluation of all
tree-level $3$-point amplitudes and the $4$-point amplitudes with
massless external states. The latter is in agreement with the results
of \cite{Leite:2016fno}.

\subsection{Cohomology}

The BRST cohomology at ghost number zero is given by the identity
operator. At ghost number one, the cohomology contains only the zero-momentum
operators $c_{+}P_{m}^{+}$ and $c_{-}P_{m}^{-}$.

Physical states are defined as elements of the BRST cohomology with
ghost number two, conformal weight zero and annihilated by the zero
mode of $b=b_{+}+b_{-}$. They can be generically expressed as $U=U_{0}+U_{+}+U_{-}$,
such that
\begin{eqnarray}
U_{0} & = & c_{+}c_{-}P_{m}^{+}P_{n}^{-}A^{mn}+\mathcal{T}(c_{+}\partial^{2}c_{+}+c_{-}\partial^{2}c_{-})A+\mathcal{T}(c_{+}\partial^{2}c_{+}-c_{-}\partial^{2}c_{-})B\nonumber \\
 &  & +c_{+}P_{m}^{+}(\partial c_{+}-\partial c_{-})A^{m}+c_{-}P_{m}^{-}(\partial c_{+}-\partial c_{-})B^{m},\label{eq:massless}\\
U_{+} & = & c_{+}c_{-}P_{m}^{+}P_{n}^{+}C_{+}^{mn}+c_{-}P_{m}^{+}(\partial c_{+}-\partial c_{-})C_{+}^{m}+c_{+}c_{-}\partial P_{m}^{+}D_{+}^{m}\nonumber \\
 &  & +\mathcal{T}c_{-}\partial^{2}c_{+}C^{+}+b_{+}c_{+}c_{-}(\partial c_{+}-\partial c_{-})D^{+},\label{eq:massive+}\\
U_{-} & = & c_{+}c_{-}P_{m}^{-}P_{n}^{-}C_{-}^{mn}+c_{+}P_{m}^{-}(\partial c_{+}-\partial c_{-})C_{-}^{m}-c_{+}c_{-}\partial P_{m}^{-}D_{-}^{m}\nonumber \\
 &  & +\mathcal{T}c_{+}\partial^{2}c_{-}C^{-}+b_{-}c_{+}c_{-}(\partial c_{+}-\partial c_{-})D^{-}.\label{eq:massive-}
\end{eqnarray}
$U_{0}$ describes massless fields ($A^{mn}$, $A^{m}$, $B^{m}$,
$A$ and $B$), while $U_{\pm}$ describes fields of mass $m^{2}=\pm4\mathcal{T}$
($C_{\pm}^{mn}$, $C_{\pm}^{m}$, $D_{\pm}^{m}$, $C^{\pm}$ and $D^{\pm}$).
The equations of motion are given by
\begin{equation}
\begin{array}{rclrcl}
\Box A^{mn} & = & 2(\partial^{n}A^{m}+\partial^{m}B^{n}), & \left(\Box\mp4\mathcal{T}\right)C_{\pm}^{mn} & = & \partial^{m}C_{\pm}^{n}+\partial^{n}C_{\pm}^{m}\mp\eta^{mn}D^{\pm},\\
\Box A & = & \partial_{m}(A^{m}+B^{m}) & \left(\Box\mp4\mathcal{T}\right)C^{\pm} & = & 2\partial_{m}C_{\pm}^{m}\mp6D^{\pm},\\
\Box B & = & \partial_{m}(A^{m}-B^{m}), & \left(\Box\mp4\mathcal{T}\right)D_{\pm}^{m} & = & 4\mathcal{T}C_{\pm}^{m}+2\partial^{m}D^{\pm},\\
A^{m} & = & \tfrac{1}{2}[\partial_{n}A^{mn}-\partial^{m}(A-B)], & C_{\pm}^{m}\pm D_{\pm}^{m} & = & \partial_{n}C_{\pm}^{mn}-\tfrac{1}{2}\partial^{m}C^{\pm},\\
B^{m} & = & \tfrac{1}{2}[\partial_{n}A^{nm}-\partial^{m}(A+B)], & D^{\pm}-\tfrac{1}{2}\partial_{m}D_{\pm}^{m} & = & \tfrac{\mathcal{T}}{2}(C_{\pm}^{mn}\eta_{mn}-3C^{\pm}),
\end{array}
\end{equation}
with gauge transformations
\begin{equation}
\begin{array}{rclrcl}
\delta A^{mn} & = & \partial^{n}\alpha^{m}+\partial^{m}\beta^{n}, & \delta C_{\pm}^{mn} & = & \partial^{m}\lambda_{\pm}^{n}+\partial^{n}\lambda_{\pm}^{m}\mp\eta^{mn}\sigma^{\pm},\\
\delta A^{m} & = & \tfrac{1}{2}\Box\alpha^{m}+\tfrac{1}{2}\partial^{m}\omega, & \delta C_{\pm}^{m} & = & \left(\Box\mp4\mathcal{T}\right)\lambda_{\pm}^{m},\\
\delta B^{m} & = & \tfrac{1}{2}\Box\beta^{m}-\tfrac{1}{2}\partial^{m}\omega, & \delta D_{\pm}^{m} & = & 2\partial^{m}\sigma^{\pm}+4\mathcal{T}\lambda_{\pm}^{m},\\
\delta A & = & \tfrac{1}{2}\partial_{m}(\alpha^{m}+\beta^{m}), & \delta C^{\pm} & = & 2\partial_{m}\lambda_{\pm}^{m}\mp6\sigma^{\pm},\\
\delta B & = & \omega+\tfrac{1}{2}\partial_{m}(\alpha^{m}-\beta^{m}), & \delta D^{\pm} & = & \left(\Box\mp4\mathcal{T}\right)\sigma^{\pm}.
\end{array}
\end{equation}
Here, $\alpha^{m}$, $\beta^{m}$, $\omega$, $\lambda_{\pm}^{m}$
and $\sigma^{\pm}$ are the gauge parameters.

In order to make the physical degrees of freedom more transparent,
it is convenient to make some field redefinitions. For the massless
vertex $U_{0}$, consider
\begin{equation}
\begin{array}{rclcrcl}
g^{m} & \equiv & A^{m}+B^{m}-\partial^{m}A, &  & b^{m} & \equiv & A^{m}-B^{m}-\partial^{m}B,\\
g^{mn} & \equiv & \tfrac{1}{2}(A^{mn}+A^{nm}), &  & b^{mn} & \equiv & \tfrac{1}{2}(A^{mn}-A^{nm}),\\
\phi & \equiv & \mathcal{T}(\tfrac{1}{2}A^{mn}\eta_{mn}-A), &  & b^{mnp} & \equiv & \tfrac{1}{3}(\partial^{p}b^{mn}+\partial^{m}b^{np}+\partial^{n}b^{pm}),
\end{array}
\end{equation}
such that the equations of motion above are rewritten as
\begin{equation}
\begin{array}{rclcrcl}
g^{m} & = & \partial_{n}g^{mn}-\eta_{np}\partial^{m}g^{np}+\tfrac{2}{\mathcal{T}}\partial^{m}\phi, &  & b^{m} & = & \partial_{n}b^{mn},\\
\Box g^{mn} & = & \partial_{p}\partial^{n}g^{mp}+\partial_{p}\partial^{m}g^{np}-\eta_{pq}\partial^{m}\partial^{n}g^{pq}+\tfrac{2}{\mathcal{T}}\partial^{m}\partial^{n}\phi, &  & \partial_{p}b^{mnp} & = & 0,\\
\Box\phi & = & 0,
\end{array}
\end{equation}
with gauge transformations
\begin{equation}
\begin{array}{rcl}
\delta\phi & = & 0,\\
\delta g^{mn} & = & \tfrac{1}{4}\partial^{n}(\alpha^{m}+\beta^{m})+\tfrac{1}{4}\partial^{m}(\alpha^{n}+\beta^{n}),\\
\delta b^{mn} & = & \tfrac{1}{4}\partial^{n}(\alpha^{m}-\beta^{m})-\tfrac{1}{4}\partial^{m}(\alpha^{n}-\beta^{n}).
\end{array}
\end{equation}
$\phi$ corresponds to the dilaton, $b^{mn}$ is the Kalb-Ramond $2$-form
and $g^{mn}$ is the graviton.

For the vertices $U_{+}$ and $U_{-}$, consider the combinations
\begin{equation}
\begin{array}{rclrcl}
d_{\pm}^{mn} & \equiv & C_{\pm}^{mn}-\tfrac{1}{4\mathcal{T}}(\partial^{n}D_{\pm}^{m}+\partial^{m}D_{\pm}^{n}) & d_{\pm}^{m} & \equiv & D_{\pm}^{m}\pm\tfrac{1}{10}(\eta_{np}\partial^{m}C_{\pm}^{np}-\partial^{m}C^{\pm}),\\
 &  & \pm\tfrac{1}{20\mathcal{T}}(\partial^{m}\partial^{n}\pm\mathcal{T}\eta^{mn})(C^{\pm}-C_{\pm}^{pq}\eta_{pq}) & d_{\pm} & \equiv & C^{\pm}-C_{\pm}^{mn}\eta_{mn},
\end{array}
\end{equation}
with gauge transformations
\begin{equation}
\begin{array}{ccc}
\delta d_{\pm}=\pm20\sigma^{\pm}, & \delta d_{\pm}^{m}=4\mathcal{T}\lambda_{\pm}^{m}, & \delta d_{\pm}^{mn}=0.\end{array}
\end{equation}
The fields $d_{\pm}$ and $d_{\pm}^{m}$ are pure gauge, while $d_{\pm}^{mn}$
corresponds to spin $2$ fields with $m^{2}=\pm4\mathcal{T}$ satisfying
\begin{equation}
\begin{array}{ccc}
\left(\Box\mp4\mathcal{T}\right)d_{\pm}^{mn}=0, & \partial_{n}d_{\pm}^{mn}=0, & d_{\pm}^{mn}\eta_{mn}=0.\end{array}
\end{equation}

\subsection{$3$-point amplitudes at tree level}

For the amplitude evaluations, it is simpler to consider a gauge fixed
form of the vertices above with momentum eigenstates. In addition,
it might be helpful to work with their sector-split versions, \emph{cf.}
definition \eqref{eq:uvobar}, in order to build some intuition on
the $3$-point amplitudes and their independence from the fixed positions
of the vertices (M\"{o}bius invariance). The gauge fixed vertices
will be chosen to be:
\begin{itemize}
\item massless vertex, $U_{0}$, with sector-split form
\begin{equation}
\bar{U}_{0}(z^{+},z^{-};k_{m})=A^{mn}c_{+}P_{m}^{+}(z^{+})c_{-}P_{n}^{-}(z^{-})e^{ik\cdot\bar{X}(z^{+},z^{-})},\label{eq:masslessgaugefixed}
\end{equation}
with $k_{m}k^{m}=0$, $k_{n}A^{mn}=k_{n}A^{nm}=0$;
\item massive vertex, $U_{+}$, with sector-split form 
\begin{equation}
\bar{U}_{+}(z^{+},z^{-};k_{m})=C_{+}^{mn}c_{+}P_{m}^{+}P_{n}^{+}(z^{+})c_{-}(z^{-})e^{ik\cdot\bar{X}(z^{+},z^{-})}
\end{equation}
with $k_{m}k^{m}=-4\mathcal{T}$, $k_{n}C_{+}^{mn}=0$ and $C_{+}^{mn}\eta_{mn}=0$;
\item and the tachyonic vertex, $U_{-}$, with sector-split form
\begin{equation}
\bar{U}_{-}(z^{+},z^{-};k_{m})=C_{-}^{mn}c_{+}(z^{+})c_{-}P_{m}^{-}P_{n}^{-}(z^{-})e^{ik\cdot\bar{X}(z^{+},z^{-})},
\end{equation}
with $k_{m}k^{m}=4\mathcal{T}$, $k_{n}C_{-}^{mn}=0$ and $C_{-}^{mn}\eta_{mn}=0$;
\end{itemize}
\

There are ten $3$-point tree level amplitudes that can be built from
$U_{0}$ and $U_{\pm}$, which will be generically cast as
\begin{equation}
\mathcal{A}_{3}=\left\langle \prod_{i=1}^{3}\bar{U}_{i}(z_{i}^{+},z_{i}^{-};k_{m}^{i})\right\rangle .
\end{equation}
The contributions from the ghost measures are simply
\begin{equation}
\left\langle c_{\pm}(z_{1}^{\pm})c_{\pm}(z_{2}^{\pm})c_{\pm}(z_{3}^{\pm})\right\rangle =z_{12}^{\pm}z_{23}^{\pm}z_{31}^{\pm},
\end{equation}
where $z_{ij}^{\pm}=z_{i}^{\pm}-z_{j}^{\pm}$. The matter contribution
is obtained with the OPE reduction of the amplitude by computing all
the contractions of the $P_{m}^{\pm}$'s with themselves and with
the exponentials $e^{ik\cdot\bar{X}}$. At the end, the remaining
contribution from the exponentials is given by equation \eqref{eq:AbarXPamplitude}.
For $3$-point amplitudes, it will be defined as
\begin{eqnarray}
I_{3}(k,z^{+},z^{-}) & \equiv & \left\langle :e^{ik^{1}\cdot\bar{X}(z_{1}^{+},z_{1}^{-})}::e^{ik^{2}\cdot\bar{X}(z_{2}^{+},z_{2}^{-})}::e^{ik^{3}\cdot\bar{X}(z_{3}^{+},z_{3}^{-})}:\right\rangle ,\nonumber \\
 & = & \delta^{26}(k^{1}+k^{2}+k^{3})\left(\frac{z_{12}^{+}}{z_{12}^{-}}\right)^{\frac{(k_{1}\cdot k_{2})}{2\mathcal{T}}}\left(\frac{z_{23}^{+}}{z_{23}^{-}}\right)^{\frac{(k_{2}\cdot k_{3})}{2\mathcal{T}}}\left(\frac{z_{31}^{+}}{z_{31}^{-}}\right)^{\frac{(k_{3}\cdot k_{1})}{2\mathcal{T}}}.\label{eq:KobaNielsen3pt}
\end{eqnarray}
Although $z_{ij}^{+}$ and $z_{ij}^{-}$ appear with opposite powers,
this is precisely the combination needed to show the M\"{o}bius invariance
of the amplitudes. The list of all $3$-point amplitudes can be cast
as:
\begin{eqnarray}
\mathcal{A}_{000} & = & A^{m\bar{m}}A^{n\bar{n}}A^{p\bar{p}}T_{mnp}^{+}T_{\bar{m}\bar{n}\bar{p}}^{-}\delta^{26}(k^{1}+k^{2}+k^{3}),\\
\mathcal{A}_{00+} & = & (k_{\bar{m}}^{3}k_{\bar{n}}^{3}+2\mathcal{T}\eta_{\bar{m}\bar{n}})D_{mnpq}^{+}A^{m\bar{m}}A^{n\bar{n}}C_{+}^{pq}\delta^{26}(k^{1}+k^{2}+k^{3}),\\
\mathcal{A}_{00-} & = & (k_{m}^{3}k_{n}^{3}-2\mathcal{T}\eta_{mn})D_{\bar{m}\bar{n}\bar{p}\bar{q}}^{-}A^{m\bar{m}}A^{n\bar{n}}C_{-}^{\bar{p}\bar{q}}\delta^{26}(k^{1}+k^{2}+k^{3}),\\
\mathcal{A}_{0++} & = & \tfrac{1}{2}(k_{\bar{m}}^{2}-k_{\bar{m}}^{3})E_{mnpqr}^{+}A^{m\bar{m}}C_{+}^{np}C_{+}^{qr}\delta^{26}(k^{1}+k^{2}+k^{3}),\\
\mathcal{A}_{0--} & = & \tfrac{1}{2}(k_{m}^{2}-k_{m}^{3})E_{\bar{m}\bar{n}\bar{p}\bar{q}\bar{r}}^{-}A^{m\bar{m}}C_{-}^{\bar{n}\bar{p}}C_{-}^{\bar{q}\bar{r}}\delta^{26}(k^{1}+k^{2}+k^{3}),\\
\mathcal{A}_{0+-} & = & F_{mnp}^{+}F_{\bar{m}\bar{n}\bar{p}}^{-}A^{m\bar{m}}C_{+}^{np}C_{-}^{\bar{n}\bar{p}}\delta^{26}(k^{1}+k^{2}+k^{3}),\\
\mathcal{A}_{++-} & = & k_{\bar{m}}^{1}k_{\bar{n}}^{2}G_{mnpq}^{+}C_{+}^{mn}C_{+}^{pq}C_{-}^{\bar{m}\bar{n}}\delta^{26}(k^{1}+k^{2}+k^{3}),\\
\mathcal{A}_{+--} & = & k_{m}^{1}k_{n}^{2}G_{\bar{m}\bar{n}\bar{p}\bar{q}}^{-}C_{-}^{\bar{m}\bar{n}}C_{-}^{\bar{p}\bar{q}}C_{+}^{mn}\delta^{26}(k^{1}+k^{2}+k^{3}),\\
\mathcal{A}_{+++} & = & H_{mnpqrs}^{+}C_{+}^{mn}C_{+}^{pq}C_{+}^{rs}\delta^{26}(k^{1}+k^{2}+k^{3}),\\
\mathcal{A}_{---} & = & H_{\bar{m}\bar{n}\bar{p}\bar{q}\bar{r}\bar{s}}^{-}C_{-}^{\bar{m}\bar{n}}C_{-}^{\bar{p}\bar{q}}C_{-}^{\bar{r}\bar{s}}\delta^{26}(k^{1}+k^{2}+k^{3}),
\end{eqnarray}
where $A_{m\bar{m}}$, $C_{mn}^{+}$ and $C_{\bar{m}\bar{n}}^{-}$
are the vertex polarizations as presented above and the kinematic
tensors are given by
\begin{eqnarray}
T_{mnp}^{\pm} & \equiv & k_{m}^{2}k_{n}^{3}k_{p}^{1}\pm2\mathcal{T}(k_{m}^{2}\eta_{np}+k_{n}^{3}\eta_{mp}+k_{p}^{1}\eta_{mn}),\\
D_{mnpq}^{\pm} & \equiv & 8\mathcal{T}^{2}\eta_{mp}\eta_{nq}\mp2\mathcal{T}(2\eta_{mp}k_{n}^{1}k_{q}^{1}+2\eta_{np}k_{m}^{2}k_{q}^{2}-\eta_{mn}k_{p}^{1}k_{q}^{2})-k_{p}^{1}k_{q}^{2}k_{m}^{3}k_{n}^{3},\\
E_{mnpqr}^{\pm} & \equiv & 4\mathcal{T}\eta_{mn}(4\mathcal{T}\eta_{pq}\mp k_{p}^{1}k_{q}^{1})k_{r}^{1}-4\mathcal{T}\eta_{mq}(4\mathcal{T}\eta_{nr}\mp k_{n}^{1}k_{r}^{1})k_{p}^{1}\nonumber \\
 &  & +\tfrac{1}{4}(k_{m}^{2}-k_{m}^{3})(4\mathcal{T}\eta_{nr}\mp k_{n}^{1}k_{r}^{1})(4\mathcal{T}\eta_{pq}\mp k_{p}^{1}k_{q}^{1})\nonumber \\
 &  & \mp\mathcal{T}(k_{m}^{2}-k_{m}^{3})(\eta_{pq}k_{n}^{1}k_{r}^{1}+\eta_{nr}k_{p}^{1}k_{q}^{1})+\tfrac{1}{4}(k_{m}^{2}-k_{m}^{3})k_{n}^{1}k_{r}^{1}k_{p}^{1}k_{q}^{1},\\
F_{mnp}^{\pm} & \equiv & k_{n}^{1}k_{p}^{1}k_{m}^{3}\pm2\mathcal{T}(\eta_{mn}k_{p}^{1}+\eta_{mp}k_{n}^{1}),\\
G_{mnpq}^{\pm} & \equiv & 4\mathcal{T}^{2}(\eta_{mp}\eta_{nq}+\eta_{mq}\eta_{np})-k_{m}^{3}k_{n}^{3}k_{p}^{3}k_{q}^{3}\nonumber \\
 &  & \mp2\mathcal{T}(\eta_{mp}k_{n}^{3}k_{q}^{3}+\eta_{mq}k_{n}^{3}k_{p}^{3}+\eta_{np}k_{m}^{3}k_{q}^{3}+\eta_{nq}k_{m}^{3}k_{p}^{3}),\\
H_{mnpqrs}^{\pm} & \equiv & k_{r}^{1}k_{s}^{1}k_{m}^{2}k_{n}^{2}k_{p}^{3}k_{q}^{3}\pm32\mathcal{T}^{3}(\eta_{mp}\eta_{nr}\eta_{qs}+\eta_{mr}\eta_{np}\eta_{qs})\nonumber \\
 &  & +8\mathcal{T}^{2}(2\eta_{mp}\eta_{nr}k_{q}^{3}+2\eta_{mr}\eta_{np}k_{q}^{3}+\eta_{mp}\eta_{nq}k_{r}^{1})k_{s}^{1}\nonumber \\
 &  & +8\mathcal{T}^{2}(2\eta_{np}\eta_{qr}k_{s}^{1}+2\eta_{nq}\eta_{pr}k_{s}^{1}+\eta_{pr}\eta_{qs}k_{n}^{2})k_{m}^{2}\nonumber \\
 &  & +8\mathcal{T}^{2}(2\eta_{nr}\eta_{ps}k_{m}^{2}+2\eta_{ns}\eta_{pr}k_{m}^{2}+\eta_{mr}\eta_{ns}k_{p}^{3})k_{q}^{3}\nonumber \\
 &  & \pm2\mathcal{T}(\eta_{pr}k_{s}^{1}k_{q}^{3}+\eta_{ps}k_{r}^{1}k_{q}^{3}+\eta_{qr}k_{s}^{1}k_{p}^{3}+\eta_{qs}k_{r}^{1}k_{p}^{3})k_{m}^{2}k_{n}^{2}\nonumber \\
 &  & \pm2\mathcal{T}(\eta_{mr}k_{s}^{1}k_{n}^{2}+\eta_{ms}k_{r}^{1}k_{n}^{2}+\eta_{nr}k_{s}^{1}k_{m}^{2}+\eta_{ns}k_{r}^{1}k_{m}^{2})k_{p}^{3}k_{q}^{3}\nonumber \\
 &  & \pm2\mathcal{T}(\eta_{mp}k_{n}^{2}k_{q}^{3}+\eta_{np}k_{m}^{2}k_{q}^{3}+\eta_{mq}k_{n}^{2}k_{p}^{3}+\eta_{nq}k_{m}^{2}k_{p}^{3})k_{r}^{1}k_{s}^{1}.
\end{eqnarray}
Note, in particular, that the $3$-point amplitude with massless vertices,
$\mathcal{A}_{000}$, is in agreement with the results of \cite{Huang:2016bdd}.

\subsection{$4$-point amplitude with massless external states}

As an example, the $4$-point amplitude with external massless states
will be now analyzed. In addition to M\"{o}bius invariance, it will
be shown that the amplitude has the expected poles located over the
mass spectrum of the physical states.

Consider
\begin{equation}
\mathcal{A}_{0000}=\left\langle \bar{U}_{0}(z_{1}^{+},z_{1}^{-};k^{1})\bar{U}_{0}(z_{2}^{+},z_{2}^{-};k^{2})\bar{U}_{0}(z_{3}^{+},z_{3}^{-};k^{3})\mathcal{V}_{0}(k^{4})\right\rangle ,
\end{equation}
where the integrated vertex operator, $\mathcal{V}_{0}(k^{4})$, is
given by 
\begin{equation}
\mathcal{V}_{0}(k^{4})=\tfrac{1}{2\mathcal{T}}A^{mn}\int_{S^{2}}dz^{+}dz^{-}P_{m}^{+}(z^{+})P_{n}^{-}(z^{-})e^{ik^{4}\cdot\bar{X}(z^{+},z^{-})},
\end{equation}
 \emph{cf.} equation \eqref{eq:ivowithintegral} and the gauge fixed
unintegrated vertex operator \eqref{eq:masslessgaugefixed}.

In this case, the Mandelstam variables are such that
\begin{equation}
\begin{array}{rclcrclcrcl}
s & \equiv & -(k^{1}+k^{2})^{2}, &  & t & \equiv & -(k^{1}+k^{3})^{2}, &  & u & \equiv & -(k^{1}+k^{4})^{2},\\
 & = & -2(k^{1}\cdot k^{2}), &  &  & = & -2(k^{1}\cdot k^{3}), &  &  & = & -2(k^{1}\cdot k^{4}),
\end{array}
\end{equation}
and $s+t+u=0$. Using the results of subsection \ref{subsec:ivoscattering},
the computation of $\mathcal{A}_{0000}$ is straightforward and can
be cast as
\begin{equation}
\mathcal{A}_{0000}=\tfrac{1}{2\mathcal{T}}A^{m\bar{m}}A^{n\bar{n}}A^{p\bar{p}}A^{q\bar{q}}\int_{S^{2}}dz_{4}^{+}dz_{4}^{-}\{T_{mnpq}^{+}\,T_{\bar{m}\bar{n}\bar{p}\bar{q}}^{-}\,I_{4}(k,z^{+},z^{-})\},\label{eq:A4g}
\end{equation}
where
\begin{eqnarray}
I_{4}(k,z^{+},z^{-}) & \equiv & \left\langle :e^{ik^{1}\cdot\bar{X}(z_{1}^{+},z_{1}^{-})}::e^{ik^{2}\cdot\bar{X}(z_{2}^{+},z_{2}^{-})}::e^{ik^{3}\cdot\bar{X}(z_{3}^{+},z_{3}^{-})}::e^{ik^{4}\cdot\bar{X}(z_{4}^{+},z_{4}^{-})}:\right\rangle ,\nonumber \\
 & = & \delta^{26}(k^{1}+k^{2}+k^{3}+k^{4})\left(\frac{z_{12}^{+}z_{34}^{+}}{z_{23}^{+}z_{14}^{+}}\frac{z_{23}^{-}z_{14}^{-}}{z_{12}^{-}z_{34}^{-}}\right)^{-\frac{s}{4\mathcal{T}}}\left(\frac{z_{31}^{+}z_{24}^{+}}{z_{23}^{+}z_{14}^{+}}\frac{z_{23}^{-}z_{14}^{-}}{z_{31}^{-}z_{24}^{-}}\right)^{-\frac{t}{4\mathcal{T}}}.\label{eq:4ptexp}
\end{eqnarray}
The dynamical tensors $T_{mnpq}^{+}$ and $T_{\bar{m}\bar{n}\bar{p}\bar{q}}^{-}$
are defined as
\begin{eqnarray}
T_{mnpq}^{\pm} & \equiv & 4\mathcal{T}^{2}\eta_{mn}\eta_{pq}\frac{z_{12}^{\pm}z_{23}^{\pm}z_{31}^{\pm}}{(z_{12}^{\pm})^{2}(z_{34}^{\pm})^{2}}+4\mathcal{T}^{2}\eta_{mp}\eta_{nq}\frac{z_{12}^{\pm}z_{23}^{\pm}z_{31}^{\pm}}{(z_{13}^{\pm})^{2}(z_{24}^{\pm})^{2}}+4\mathcal{T}^{2}\eta_{mq}\eta_{np}\frac{z_{12}^{\pm}z_{23}^{\pm}z_{31}^{\pm}}{(z_{14}^{\pm})^{2}(z_{23}^{\pm})^{2}}\nonumber \\
 &  & \pm2\mathcal{T}\eta_{mq}\frac{z_{12}^{\pm}z_{23}^{\pm}z_{31}^{\pm}}{(z_{14}^{\pm})^{2}}V_{n}^{2}V_{p}^{3}\pm2\mathcal{T}\eta_{nq}\frac{z_{12}^{\pm}z_{23}^{\pm}z_{31}^{\pm}}{(z_{24}^{\pm})^{2}}V_{m}^{1}V_{p}^{3}\pm2\mathcal{T}\eta_{pq}\frac{z_{12}^{\pm}z_{23}^{\pm}z_{31}^{\pm}}{(z_{34}^{\pm})^{2}}V_{m}^{1}V_{n}^{2}\nonumber \\
 &  & \pm2\mathcal{T}\eta_{mn}\frac{z_{12}^{\pm}z_{23}^{\pm}z_{31}^{\pm}}{(z_{12}^{\pm})^{2}}V_{p}^{3}V_{q}^{4}\pm2\mathcal{T}\eta_{mp}\frac{z_{12}^{\pm}z_{23}^{\pm}z_{31}^{\pm}}{(z_{13}^{\pm})^{2}}V_{n}^{2}V_{q}^{4}\pm2\mathcal{T}\eta_{np}\frac{z_{12}^{\pm}z_{23}^{\pm}z_{31}^{\pm}}{(z_{23}^{\pm})^{2}}V_{m}^{1}V_{q}^{4}\nonumber \\
 &  & +z_{12}^{\pm}z_{23}^{\pm}z_{31}^{\pm}V_{m}^{1}V_{n}^{2}V_{p}^{3}V_{q}^{4},
\end{eqnarray}
with
\begin{equation}
\begin{array}{rclcrcl}
V_{m}^{1} & \equiv & k_{m}^{2}\frac{z_{24}^{\pm}}{z_{12}^{\pm}z_{14}^{\pm}}+k_{m}^{3}\frac{z_{34}^{\pm}}{z_{13}^{\pm}z_{14}^{\pm}}, &  & V_{q}^{4} & \equiv & \tfrac{1}{3}k_{q}^{1}\left(\frac{z_{12}^{\pm}}{z_{14}^{\pm}z_{24}^{\pm}}-\frac{z_{31}^{\pm}}{z_{14}^{\pm}z_{34}^{\pm}}\right)\\
V_{n}^{2} & \equiv & k_{n}^{1}\frac{z_{14}^{\pm}}{z_{21}^{\pm}z_{24}^{\pm}}+k_{n}^{3}\frac{z_{34}^{\pm}}{z_{23}^{\pm}z_{24}^{\pm}}, &  &  &  & +\tfrac{1}{3}k_{q}^{2}\left(\frac{z_{23}^{\pm}}{z_{24}^{\pm}z_{34}^{\pm}}-\frac{z_{12}^{\pm}}{z_{14}^{\pm}z_{24}^{\pm}}\right)\\
V_{p}^{3} & \equiv & k_{p}^{1}\frac{z_{14}^{\pm}}{z_{31}^{\pm}z_{34}^{\pm}}+k_{p}^{2}\frac{z_{24}^{\pm}}{z_{32}^{\pm}z_{34}^{\pm}}, &  &  &  & +\tfrac{1}{3}k_{q}^{3}\left(\frac{z_{31}^{\pm}}{z_{14}^{\pm}z_{34}^{\pm}}-\frac{z_{23}^{\pm}}{z_{24}^{\pm}z_{34}^{\pm}}\right).
\end{array}
\end{equation}

The amplitude $\mathcal{A}_{0000}$ is independent of the position
of the unintegrated vertices. The M\"{o}bius transformations are
given by
\begin{equation}
\begin{array}{cc}
z_{i}^{+}\to\frac{az_{i}^{+}+b}{cz_{i}^{+}+d}, & z_{i}^{-}\to\frac{a^{*}z_{i}^{-}+b^{*}}{c^{*}z_{i}^{-}+d^{*}},\end{array}
\end{equation}
with $ad-bc=1.$ Therefore, $z_{ij}^{\pm}$ transforms as
\begin{equation}
\begin{array}{cc}
z_{ij}^{+}\to\frac{z_{ij}^{+}}{(cz_{i}^{+}+d)(cz_{j}^{+}+d)}, & z_{ij}^{-}\to\frac{z_{ij}^{-}}{(c^{*}z_{i}^{-}+d^{*})(c^{*}z_{j}^{-}+d^{*})},\end{array}\label{eq:mobiuszij}
\end{equation}
naturally leading to the definition of the invariant cross-ratios
\begin{equation}
\begin{array}{ccc}
x^{\pm}\equiv\frac{z_{41}^{\pm}z_{23}^{\pm}}{z_{12}^{\pm}z_{34}^{\pm}}, & 1-x^{\pm}=\frac{z_{13}^{\pm}z_{24}^{\pm}}{z_{12}^{\pm}z_{34}^{\pm}}, & \tfrac{x^{\pm}}{1-x^{\pm}}=\frac{z_{41}^{\pm}z_{23}^{\pm}}{z_{13}^{\pm}z_{24}^{\pm}}.\end{array}
\end{equation}
In terms of $x^{\pm}$, equation \eqref{eq:4ptexp} can be expressed
as
\begin{equation}
I_{4}(s,u;x^{\pm})=\delta^{26}(k^{1}+k^{2}+k^{3}+k^{4})\left(\frac{x^{-}}{x^{+}}\right)^{\frac{u}{4\mathcal{T}}}\left(\frac{1-x^{-}}{1-x^{+}}\right)^{\frac{t}{4\mathcal{T}}},
\end{equation}
and the M\"{o}bius invariance of $\mathcal{A}_{0000}$ can be easily
checked using the transformations \eqref{eq:mobiuszij} and
\begin{equation}
dz_{4}^{+}dz_{4}^{-}\to\frac{dz_{4}^{+}dz_{4}^{-}}{(cz_{4}^{+}+d)^{2}(c^{*}z_{4}^{-}+d^{*})^{2}}.
\end{equation}

Therefore, the positions $z_{1}^{\pm}$, $z_{2}^{\pm}$ and $z_{3}^{\pm}$
will be conveniently chosen to be $z_{1}^{\pm}=0$, $z_{2}^{\pm}=1$
and $z_{3}^{\pm}=\infty$, such that $x^{\pm}=z_{4}^{\pm}$ and

\begin{eqnarray}
T_{mnpq}^{\pm} & = & -\frac{(k_{m}^{2}x^{\pm}+k_{m}^{4})(k_{n}^{1}x^{\pm}+k_{n}^{3})(k_{p}^{4}x^{\pm}+k_{p}^{2})(k_{q}^{3}x^{\pm}+k_{q}^{1})}{(x^{\pm})^{2}(1-x^{\pm})^{2}}\nonumber \\
 &  & \mp2\mathcal{T}\eta_{nq}\frac{(k_{m}^{2}x^{\pm}+k_{m}^{4})(k_{p}^{4}x^{\pm}+k_{p}^{2})}{(x^{\pm})(1-x^{\pm})^{2}}\pm2\mathcal{T}\eta_{mp}\frac{(k_{n}^{1}x^{\pm}+k_{n}^{3})(k_{q}^{3}x^{\pm}+k_{q}^{1})}{(x^{\pm})(1-x^{\pm})^{2}}\nonumber \\
 &  & \mp2\mathcal{T}\eta_{mq}\frac{(k_{n}^{1}x^{\pm}+k_{n}^{3})(k_{p}^{4}x^{\pm}+k_{p}^{2})}{(x^{\pm})^{2}(1-x^{\pm})}\mp2\mathcal{T}\eta_{np}\frac{(k_{m}^{2}x^{\pm}+k_{m}^{4})(k_{q}^{3}x^{\pm}+k_{q}^{1})}{(x^{\pm})^{2}(1-x^{\pm})}\nonumber \\
 &  & \pm2\mathcal{T}\eta_{pq}\frac{(k_{m}^{2}x^{\pm}+k_{m}^{4})(k_{n}^{1}x^{\pm}+k_{n}^{3})}{x^{\pm}(1-x^{\pm})}\pm2\mathcal{T}\eta_{mn}\frac{(k_{p}^{4}x^{\pm}+k_{p}^{2})(k_{q}^{3}x^{\pm}+k_{q}^{1})}{(x^{\pm})(1-x^{\pm})}\nonumber \\
 &  & +4\mathcal{T}^{2}\eta_{mn}\eta_{pq}+4\mathcal{T}^{2}\frac{\eta_{mp}\eta_{nq}}{(1-x^{\pm})^{2}}+4\mathcal{T}^{2}\frac{\eta_{mq}\eta_{np}}{(x^{\pm})^{2}}.
\end{eqnarray}

Now, observe that all terms of the integrand of \eqref{eq:A4g} can
be written as
\begin{equation}
I_{4}(U,T;m,n,\bar{m},\bar{n})\equiv\int d^{2}z\left\{ z^{m-2}(1-z)^{n-2}\bar{z}^{\bar{m}-2}(1-\bar{z})^{\bar{n}-2}\left(\frac{\bar{z}}{z}\right)^{U}\left(\frac{1-\bar{z}}{1-z}\right)^{T}\right\} ,\label{eq:KLT-sectorized}
\end{equation}
where $\{m,n\}=0,\ldots,4$, with $m+n\leq4$ and the same for $\{\bar{m},\bar{n}$\}.
This restriction appears only in the chiral model and is a consequence
of the conformal weights of the building blocks of the vertex operators.
The normalized Mandelstam variables $S$, $T$ and $U$, with $S+T+U=0$,
are just
\begin{equation}
\begin{array}{ccc}
S\equiv\tfrac{s}{4\mathcal{T}}, & T\equiv\tfrac{u}{4\mathcal{T}}, & U\equiv\tfrac{u}{4\mathcal{T}}.\end{array}
\end{equation}

The chiral string character of this amplitude is very clear for the
powers involving the Mandelstam variables appear as $\tfrac{\bar{z}}{z}$
and $\tfrac{1-\bar{z}}{1-z}$ instead of $|z|^{2}$ and $|1-z|^{2}$.
The evaluation of $I_{4}$ will be done by adapting the known Kawai-Lewellen-Tye
relations of \cite{Kawai:1985xq} to the case above, which is slightly
different due to the opposite phase contribution of the branch cuts.

Considering $z=x+iy$, $I_{4}$ can be seem as an analytic function
in $y$ with four branch points,
\begin{equation}
y=\pm ix,\pm i(1-x).
\end{equation}
The contour of integration of $y$ over the real line can be deformed
to the imaginary line. To do that, it is useful to rewrite part of
the integrand as 
\begin{equation}
\left(\frac{\bar{z}}{z}\right)^{U}=\bar{z}^{2U}\frac{1}{\Gamma(U)}\int_{0}^{\infty}\alpha^{s-1}e^{-(Z\bar{Z})\alpha}d\alpha.
\end{equation}
For a fixed $\alpha$, the exponential acts as a convergence factor
in the integration and the contour deformation is well defined. The
overall result is similar to a Wick-rotation. After defining $\xi\equiv x+iy$
and $\eta\equiv x-iy$, $I_{4}$ can be written as
\begin{equation}
I_{4}=\int_{-\infty}^{+\infty}d\xi\int_{-\infty}^{+\infty}d\eta\left\{ |\xi|^{m-2-U}|1-\xi|^{n-2-T}|\eta|^{\bar{m}-2+U}|1-\eta|^{\bar{n}-2+T}\right\} \times f(\xi,\eta),
\end{equation}
where $f(\xi,\eta)$ is a phase factor determined by the domain analysis
of the integrals. Following closely the steps in \cite{Kawai:1985xq},
it is possible to show that
\begin{equation}
I_{4}=\sin(\pi T)\times\int_{0}^{1}d\xi\{\xi^{m-2-U}(1-\xi)^{n-2-T}\}\times\int_{1}^{\infty}d\eta\{\eta^{\bar{m}-2+U}(\eta-1)^{\bar{n}-2+T}\}.
\end{equation}

Now, using Euler's reflection formula,
\begin{equation}
\sin\pi a=\frac{\pi}{\Gamma(1-a)\Gamma(a)},
\end{equation}
where $\Gamma(a)$ is the gamma function, and recalling that
\begin{equation}
\int_{0}^{1}dxx^{a-1}(1-x)^{b-1}=\frac{\Gamma(a)\Gamma(b)}{\Gamma(a+b)},
\end{equation}
 $I_{4}$ is finally written (up to a sign) as
\begin{equation}
I_{4}(U,T;m,n,\bar{m},\bar{n})=\pi\frac{\Gamma(3-\bar{m}-\bar{n}+S)}{\Gamma(m+n-2+S)}\frac{\Gamma(m-1-U)}{\Gamma(2-\bar{m}-U)}\frac{\Gamma(n-1-T)}{\Gamma(2-\bar{n}-T)}.
\end{equation}
Given that $m,n,\bar{m},\bar{n}\geq0$ with $m+n\leq4$ and $\bar{m}+\bar{n}\leq4$,
it is straightforward to see that the poles of $I_{4}$ are located
at $S,T,U=0,\pm1$ (or $s,t,u=0,\pm4\mathcal{T}$), \emph{i.e.} the
expected poles associated to the physical spectrum of the model. Note,
for example, that
\begin{equation}
\frac{\Gamma(m-1-U)}{\Gamma(2-\bar{m}-U)}=\begin{cases}
{\displaystyle \prod_{i=1}^{m+\bar{m}-3}}(m-1-U-i) & m+\bar{m}\geq3,\\
{\displaystyle \prod_{i=1}^{3-m-\bar{m}}}(2-\bar{m}-U-i)^{-1} & m+\bar{m}<3,
\end{cases}
\end{equation}
and the poles occur in the $U$ channel only when $m+\bar{m}<3$.
In the $T$ channel, the poles appear when $n+\bar{n}<3$. For the
$S$ channel the poles occur when $m+n+\bar{m}+\bar{n}>5$:
\begin{equation}
\frac{\Gamma(3-\bar{m}-\bar{n}+S)}{\Gamma(m+n-2+S)}=\begin{cases}
{\displaystyle \prod_{i=1}^{5-m-n-\bar{m}-\bar{n}}}(3-\bar{m}-\bar{n}+S-i) & m+n+\bar{m}+\bar{n}\leq5,\\
{\displaystyle \prod_{i=1}^{m+n+\bar{m}+\bar{n}-5}}(m+n-2+S-i)^{-1} & m+n+\bar{m}+\bar{n}>5.
\end{cases}
\end{equation}
The cancellation of the other poles of the gamma functions works exactly
as described in \cite{Huang:2016bdd}.

\end{document}